\documentclass[useAMS,usenatbib,fleqn,twocolumn]{aastex63}
\usepackage{mathtools}
\usepackage{amssymb,amsmath}
\begin{document}

\shorttitle{Shining Wanderers}
\title{Unveiling the Population of Wandering Black Holes via Electromagnetic Signatures
}

\shortauthors{Ricarte et al.}
\correspondingauthor{Angelo Ricarte}
\email{angelo.ricarte@cfa.harvard.edu}

\author{Angelo Ricarte}
\affiliation{Center for Astrophysics | Harvard \& Smithsonian, 60 Garden Street, Cambridge, MA 02138, USA}
\affiliation{Black Hole Initiative, 20 Garden Street, Cambridge, MA 02138, USA}

\author{Michael Tremmel}
\affiliation{Department of Astronomy, Yale University, 52 Hillhouse Avenue, New Haven, CT 06511, USA}

\author{Priyamvada Natarajan}
\affiliation{Black Hole Initiative, 20 Garden Street, Cambridge, MA 02138, USA}
\affiliation{Department of Astronomy, Yale University, 52 Hillhouse Avenue, New Haven, CT 06511, USA}
\affiliation{Department of Physics, Yale University, P.O. Box 208121, New Haven, CT 06520, USA}

\author{Thomas Quinn}
\affiliation{Department of Astronomy, University of Washington, PO Box 351580, Seattle, WA 98195, USA}

\date{\today}

\begin{abstract}
While most galaxies appear to host a central supermassive black hole (SMBH), they are expected to also contain a substantial population of off-center ``wandering'' SMBHs naturally produced by the hierarchical merger-driven process of galaxy assembly.  This population has been recently characterized in an analysis of the {\sc Romulus} cosmological simulations, which correct for the dynamical forces on SMBHs without artificially pinning them to halo centers.  Here we predict an array of electromagnetic signatures for these wanderers.  The predicted wandering population of SMBHs from {\sc Romulus} broadly reproduces the observed spatial offsets of a recent sample of hyperluminous X-ray sources.  We predict that the sources with the most extreme offsets are likely to arise from SMBHs within satellite galaxies.  These simulations also predict a significant population of secondary active galactic nuclei (AGN) with luminosities at least 10\% that of the central AGN. The majority of galaxies at $z=4$ that host a central AGN with bolometric luminosity $L_\mathrm{bol}>10^{42} \ \mathrm{erg} \; \mathrm{s}^{-1}$ are predicted to host a companion off-center AGN of comparable brightness.  We demonstrate that stacked X-ray observations of similar mass galaxies may reveal a halo of collective emission attributable to these wanderers.  Finally, because wanderers dominate the population of SMBHs with masses of $\lesssim 10^7\,M_{\odot}$ in {\sc Romulus}, they may dominate tidal disruption event (TDE) rates at these masses if they retain a stellar component (e.g. a nuclear star cluster).  This could warrant an order of magnitude correction to current theoretically estimated TDE rates at low SMBH masses.
\end{abstract}


\section{Introduction}

Massive galaxies are believed to host central supermassive black holes (SMBHs), which are important for regulating gas cooling via active galactic nucleus (AGN) feedback \citep[e.g.,][]{Kormendy&Ho2013}.  When one galaxy merges with another, its central SMBH begins a journey across many orders of magnitude in spatial scale that can eventually lead to a SMBH merger \citep{Begelman+1980}. First, dynamical friction grinds down SMBH orbits on kiloparsec scales \citep{Chandrasekhar1943}.  When dynamical friction becomes less efficient on sub-kiloparsec scales, a variety of phenomena such as spherical asymmetry or the presence of gas are proposed to bridge the ``final parsec problem'' \citep[e.g.,][]{Armitage+PN2002,Holley-Bockelmann&Khan2015}.  Finally, gravitational wave emission takes over to shrink the binary orbit and drive sufficiently tight binaries together until they merge.  To make robust predictions for gravitational wave events detectable by pulsar timing arrays \citep{Hobbs+2010,NANOGrav2020} or the upcoming Laser Interferometer Space Antenna (LISA) \citep{LISA+2017} mission, it is essential to understand all steps of this complex process.

Cosmological simulations, wherein SMBHs and galaxies self-consistently evolve in near realistic environments, are excellent tools for addressing the first part of this journey.  Many numerical studies have revealed that significant delays can occur at the dynamical friction step \citep[e.g.,][]{Holley-Bockelmann+2010, Volonteri+2016,Tamfal+2018,Tremmel+2018b,Bellovary+2021}.  This is especially true in clumpy, high-redshift galaxies, which may make it difficult for offset SMBHs to reach galactic centers \citep{Biernacki+2017,Tremmel+2018b,Pfister+2019,Bortolas+2020,Ma+2021}.  We term any SMBH that is spatially offset from its galactic center a ``wandering'' SMBH.  

The {\sc Romulus} cosmological simulations are ideally suited for studying this particular problem due to its careful treatment of SMBH dynamics, which correct for dynamical friction errors that naturally arise due to gravitational softening and limited mass resolution \citep{Tremmel+2015}.  This is in contrast with most cosmological simulations, which artificially and unphysically force SMBHs to be at the centers of their host galaxies.  Some exceptions include {\sc Magneticum} \citep{Hirschmann+2014,Steinborn+2016}, {\sc Horizon-AGN} \citep{Dubois+2014,Bartlett+2021}, and recent simulations by \citet{Chen+2021}.  The larger gravitational softening length of {\sc Magneticum} results in SMBHs artificially displaced by several kpc.  Meanwhile, {\sc Horizon-AGN} includes dynamical friction from gas, but not dark matter and stars, whereas the reverse is true for {\sc Romulus}.

Previous work with the {\sc Romulus} simulations has shown that SMBHs can spend Gyrs offset from the galactic center following a galaxy merger \citep{Tremmel+2018b}. Many SMBHs never make it to the centers of their galaxies, leading to a suppressed SMBH merger rate \citep{Barausse+2020}. This results in a population of wandering SMBHs, as studied in Milky Way-like galaxies at $z=0$ in \citet{Tremmel+2018a}.  {\sc Romulus} successfully reproduces observed stellar mass-halo mass and SMBH mass-stellar mass relations from dwarf to galaxy cluster scales. The simulations also produce cosmological star formation and SMBH growth histories that are consistent with observations, as well as realistic properties for the circumgalactic and intracluster media \citep{Tremmel+2017,Ricarte+2019,Butsky+2019,Tremmel+2019,Sanchez+2019,Chadayammuri+2020}. These simulations are among the highest resolution of their class, resolving dwarf galaxies as small as $10^{7}$ M$_{\odot}$ in stellar mass with hundreds of star particles and tens of thousands of dark matter particles. This high resolution allows SMBHs to be seeded within low-mass galaxies at $z>5$ based solely on local gas properties and with no \textit{a priori} assumptions about halo occupation \citep{Tremmel+2017}. This results in a black hole occupation fraction that is purely a prediction of the simulation, and which correctly matches the current best estimates from observations \citep{Ricarte+2021}. Thanks to this albeit simple seeding prescription, SMBH evolution can be tracked within low-mass, high-redshift galaxies, which are important contributors to overall SMBH merger rates \citep{Volonteri+2020}, as well as long-lived, wandering SMBHs \citep{Tremmel+2018a,Tremmel+2018b}. To our knowledge, there is no other published cosmological simulation which seeds SMBHs based on local gas properties in a way that matches the observationally estimated SMBH occupation fraction and also allows SMBHs to wander their hosts with the addition of a dynamical friction correction.

We recently studied the broad demographics of the wandering population in {\sc Romulus}, and found they exist in numbers proportional to the host halo mass, were distributed widely throughout the halo, and tended to stay near their initial seed mass in the simulation \citep{Ricarte+2021}. In this work, we present observational predictions for detecting this population. Although previous work with these simulations concluded that detecting wanderers using razor-thin lensing arcs is infeasible \citep{Banik+2019}, we find that wanderers in {\sc Romulus} can manifest as hyperluminous X-ray sources (HLXs) \citep[e.g.,][]{King&Dehnen2005,Barrows+2019}, spatially-offset AGN \citep[e.g.,][]{Comerford+2009,Koss+2012,Mezcua+2020,Reines+2020}, and a faint halo of potentially detectable emission around a typical galaxy.  If wanderers retain their stellar counterparts, they may also produce off-center tidal disruption events (TDEs).

\section{Methods}

Here, we briefly summarize the most important aspects of the SMBH physics in the {\sc Romulus} simulations.  Additional details can be found in \citet{Tremmel+2017,Tremmel+2019}, where these simulations are introduced and described in detail.  The {\sc Romulus} simulations consist of {\sc Romulus25}, a $(25 \ \mathrm{Mpc})^3$ volume, and {\sc RomulusC}, a zoom-in simulation of a $10^{14} \ \mathrm{M}_\odot$ galaxy cluster, both run under a cosmology with $\Omega_0=0.3086$, $\Lambda=0.6914$, h$=0.6777$, $\sigma_8=0.8288$.  Both simulations have a mass resolution of $3.39 \times 10^5 $M$_{\odot}$ and $2.12 \times 10^5$M$_{\odot}$ for dark matter and gas respectively and employ a spline gravitational softening length of 350 pc (equivalent to $\sim250$ pc plummer softening). Unlike most other currently available simulations of this scale, SMBHs are seeded based on local gas properties, are not explicitly tethered to the centers of their host galaxies, and grow in a way that accounts for the resolved nearby gas kinematics.

\subsection{SMBH Dynamics and Mergers}
\label{sec:dynamics}

Central to this work is the fact that SMBH dynamical evolution can be accurately tracked down to sub-kpc scales in this simulation suite. This is done via a sub-grid correction that accounts for unresolved dynamical friction from stars and dark matter \citep{Tremmel+2015}. This frictional force is estimated by assuming a locally isotropic velocity distribution and integrating Chandrasekhar's equation \citep{Chandrasekhar1943} from the 90-degree deflection radius, $r_{90}$, out to the gravitational softening length, $\epsilon_g$ \footnote{Within this region, gravity is artificially reduced in order to avoid unrealistic, collisional behavior.}, of the SMBH. This results in an acceleration applied to each SMBH particle in the simulation given by,

\begin{equation}
    \textbf{a}_\mathrm{DF} = -4\pi G^2\rho(\mathrm{v}<\mathrm{v}_\mathrm{BH})\ln\Lambda \frac{\mathrm{\textbf{v}}_{\mathrm{BH}}}{\mathrm{v}^2_{\mathrm{BH}}},
\end{equation}

\noindent where $\ln\Lambda = \ln(\frac{\epsilon_{g}}{r_{90}})$, $\mathrm{v}_{\mathrm{BH}}$ is the SMBH's velocity relative to the local center of mass velocity of the closest 64 star and dark matter particles, $\rho$ is the mass density, and $G$ is the gravitational constant. This correction, combined with the high dark matter and stellar mass resolution in the simulations, results in realistic sinking timescales for off-center SMBHs \citep{Tremmel+2015,Tremmel+2018b}.

We adopt the SMBH merger criteria derived by \citet{Bellovary+2010}, in which two SMBHs merge if they approach within 0.7 kpc of one another (or, twice the gravitational softening length of 350 pc) and are also mutually bound ($\frac{1}{2}\Delta\textbf{v} < \Delta \textbf{a} \cdot \Delta\textbf{r}$). The limit of 0.7 kpc is justified because at separations smaller than this distance we no longer trust the simulation to correctly model the dynamics of the system, as the gravitational force between the black holes becomes increasingly unresolved. When a merger occurs, the masses of the two SMBHs are added together and the resulting SMBH is given a position and velocity such that momentum is conserved (neglecting potential recoil due to gravitational wave emission).

\subsection{SMBH Seeding}
\label{sec:seeding}

SMBHs are seeded in gas that has reached high densities (3 $m_p\; \mathrm{cm}^{-3}$, 15 times the threshold for star formation in {\sc Romulus}) while being free of metals and still at relatively high temperatures (9500-10000 K).  This method selects gas with high Jeans mass that is also collapsing quicker than the typical timescale on which such a dense gas particle will form a star ($\sim10^6$ yr). This is meant to roughly approximate the formation processes suggested by massive initial seeding models including direct collapse \citep{Lodato&Natarajan2007,Alexander&Natarajan2014,Natarajan2021} and results in SMBHs that are seeded preferentially at $z>5$ within low-mass galaxies \citep{Tremmel+2017}. \citet{Ricarte+2021} show that the resulting $z=0$ SMBH occupation fraction matches estimates based on current observations down to dwarf galaxy scales.

The initial mass of a SMBH is set to $10^6$ M$_{\odot}$, with each SMBH consuming nearby gas particles to account for their mass. This mass is somewhat higher than what is typically assumed even for the most massive direct collapse formation scenarios, but it is required that the mass be at least a few times larger than the background dark matter and star particles to avoid spurious scattering events \citep{Tremmel+2015}. It is possible that such large masses affect the growth history of these SMBH, but \citet{Ricarte+2019} show that the SMBH growth depends mostly on the galaxy properties, rather than the SMBH mass.  It may also affect their dynamical evolution, but \citet{Tremmel+2018a} show that even with their relatively large masses, the wandering SMBH population has dynamical friction sinking timescales much longer than a Hubble time in $z=0$, MW-mass halos.

\subsection{SMBH Growth and Feedback}
\label{sec:fueling}

A SMBH's accretion rate is estimated according to a Bondi-Hoyle-Lyttleton prescription \citep{Bondi1952}, modified to account for angular momentum support.  Specifically, the accretion rate is given by

\begin{equation}
    \dot{M}_\bullet = \alpha(n) 
    \begin{dcases*}
    \frac{\pi(GM_\bullet)^2\rho}{(v^2_\mathrm{bulk} + c^2_s)^{3/2}} & \text{if $v_\mathrm{bulk} > v_\theta$} \\ 
    \frac{\pi(GM_\bullet)^2\rho c_s}{(v^2_\theta + c^2_s)^{2}} & \text{if $v_\mathrm{bulk} < v_\theta$}, \\ 
    \end{dcases*}
\end{equation}

\noindent where $G$ is the gravitational constant, $M_\bullet$ is the SMBH mass, $\rho$ is the ambient mass density, $c_s$ is the ambient sound speed, $v_\theta$ is the local rotational velocity of surrounding gas, and $v_\mathrm{bulk}$ is the bulk velocity relative to the SMBH.  The coefficient $\alpha(n)$ provides a boost to the accretion rate given by

\begin{equation}
    \alpha(n) = 
    \begin{dcases*}
    \left( \frac{n}{n_{\mathrm{th},*}} \right)^2 & \text{if $n \geq n_{\mathrm{th},*}$} \\ 
    1 & \text{if $n < n_{\mathrm{th},*}$} \\ 
    \end{dcases*}
\end{equation}

\noindent where $n_{\mathrm{th},*}$ is the star formation number density threshold, below which the simulation no longer resolves the multi-phase interstellar medium, as in \citet{Booth+Schaye2009}.  During the accretion process, thermal energy is injected isotropically into the surrounding gas particles, assuming a radiative efficiency of 10\% and a feedback coupling efficiency of 2\%.  We adopt the same radiative efficiency when computing bolometric luminosities, such that $L_\mathrm{bol} = 0.1 \dot{M}_\bullet c^2$.

\subsection{Selection of Dark Matter Halos and Wandering SMBHs}
\label{sec:halos}

We follow the post-processing performed in \citet{Ricarte+2021}, using the {\sc Amiga} halo finder to identify structures \citep{Knollmann&Knebe2009}, {\sc pynbody} for particle data analysis \citep{Pontzen+2013}, and {\sc tangos} for the construction of a database of galaxy and SMBH properties \citep{Pontzen&Tremmel2018}.  As in this previous study, we designate a SMBH as a ``wanderer'' if it is further than 0.7 kpc (twice the gravitational softening length) from the center of its host halo, which is identified using a shrinking spheres method \citep{Power+2003}.  This distance threshold is used because the center of the galaxy/halo becomes ill-defined within 0.7 kpc at the gravitational force resolution of {\sc Romulus}. Further, we want to avoid selecting wandering SMBHs that may be about to merge with another SMBH (see \S\ref{sec:dynamics} above).

\section{Results}

\begin{figure*}
  \centering
  \includegraphics[width=\textwidth]{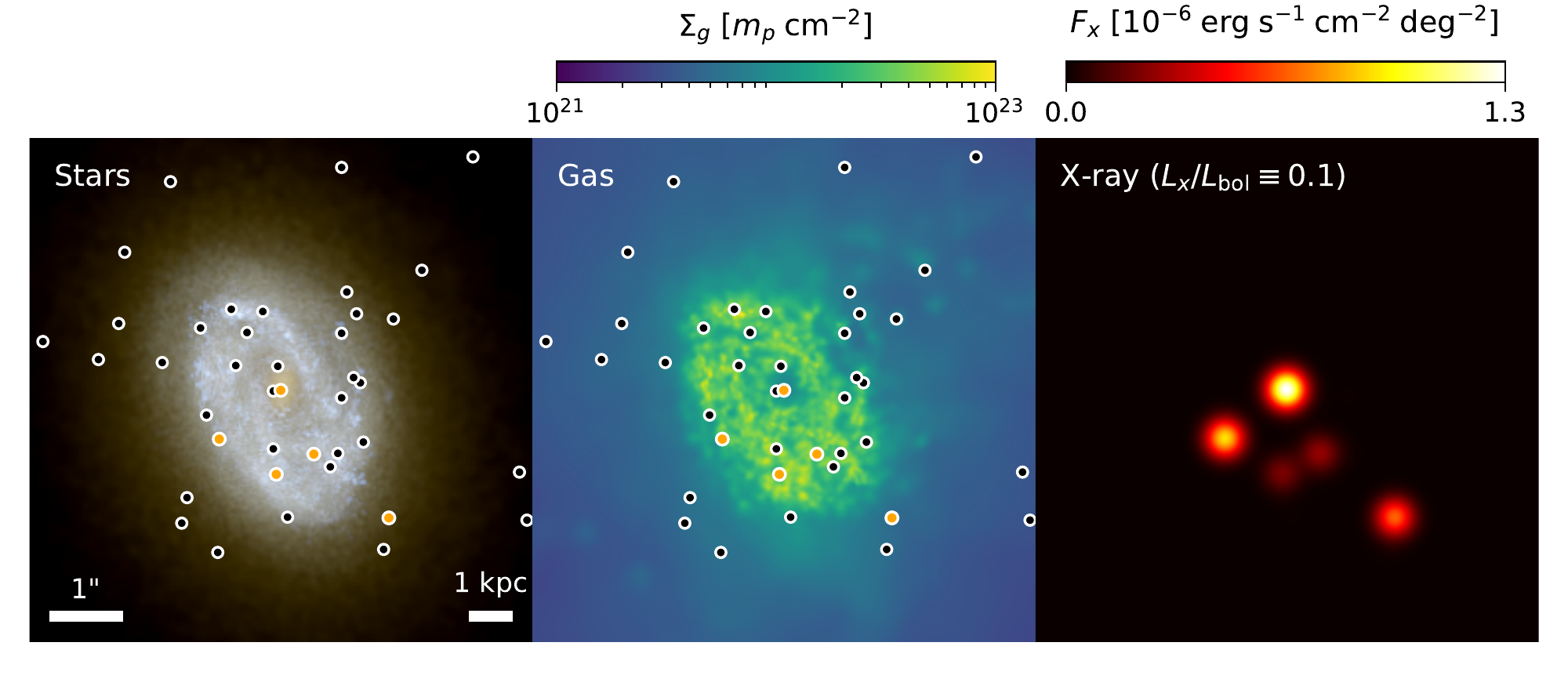}
  \caption{Example of a remarkable galaxy at $z=0.1$ which hosts 5 SMBHs each shining with a bolometric luminosity above $10^{42} \ \mathrm{erg}\;\mathrm{s}^{-1}$.  Black and orange circles demarcate the projected positions of SMBHs bound to this galaxy's halo, with orange circles marking those with $L_\mathrm{bol} >10^{42} \ \mathrm{erg} \; \mathrm{s}^{-1}$.  The brightest SMBH (at the center) shines at $L_\mathrm{bol} = 3.7 \times 10^{43} \ \mathrm{erg} \; \mathrm{s}^{-1}$.  From left to right, we plot an image of stars with a surface brightness limit of 24 $\mathrm{mag}\;\mathrm{arcsec}^{-2}$, the integrated gas surface density, and a simulated X-ray image.  In the X-ray image, we assume that 10\% of the bolometric luminosity is emitted in some X-ray band, and that the observing instrument has a point spread function with a full width at half maximum of 0.5 arcseconds, appropriate for the center of the {\it Chandra} field of view.  Despite the large number of accreting sources, there is no obvious morphological sign of a merger.  \label{fig:example}}
\end{figure*}

In the {\sc Romulus} simulations, SMBHs accrete according to the Bondi formula regardless of their positions in the halo.  For many galaxies, this may result in a very faint halo of emission, further explored in \S\ref{sec:profile}.  In Figure \ref{fig:example}, we plot a more striking example of a galaxy hosting 5 SMBHs each with bolometric luminosities $L_\mathrm{bol} >10^{42} \; \mathrm{erg}\; \mathrm{s}^{-1}$ when accretion rates are averaged over 30 Myr.  For this luminosity threshold, this galaxy hosts the most AGN at this snapshot, $z=0.1$.  The brightest SMBHs are simply the most massive ones in the halo, each SMBH having grown to a mass of $10^7 \ \mathrm{M}_\odot$ or higher, while the median wanderer mass is near the seed mass of $10^6 \ \mathrm{M}_\odot$.  In the right panel, we simulate an X-ray image by assuming that 10\% of the bolometric luminosity is emitted in some X-ray band \citep[e.g.,][]{Hopkins+2007}, and that the instrument has a point spread function with a full-width at half maximum of 0.5 arcseconds, appropriate for {\it Chandra}.  As shown in \citep{Tremmel+2018b}, SMBHs can remain offset in the {\sc Romulus} simulations Gyrs after the halo mergers which deposited them occurred.  We emphasize that the wandering SMBHs considered in this study are not necessarily associated with active galaxy mergers.  Despite the large number of accreting SMBHs in Figure \ref{fig:example}, the host galaxy morphology resembles an ordinary spiral with a stellar mass of $1.0\times 10^{11} \ \mathrm{M}_\odot$ and no obvious merger signatures.  

\subsection{Comparison to Hyperluminous X-ray (HLX) Sources}
\label{sec:hlxs}

\begin{figure*}
  \centering
  \includegraphics[width=\textwidth]{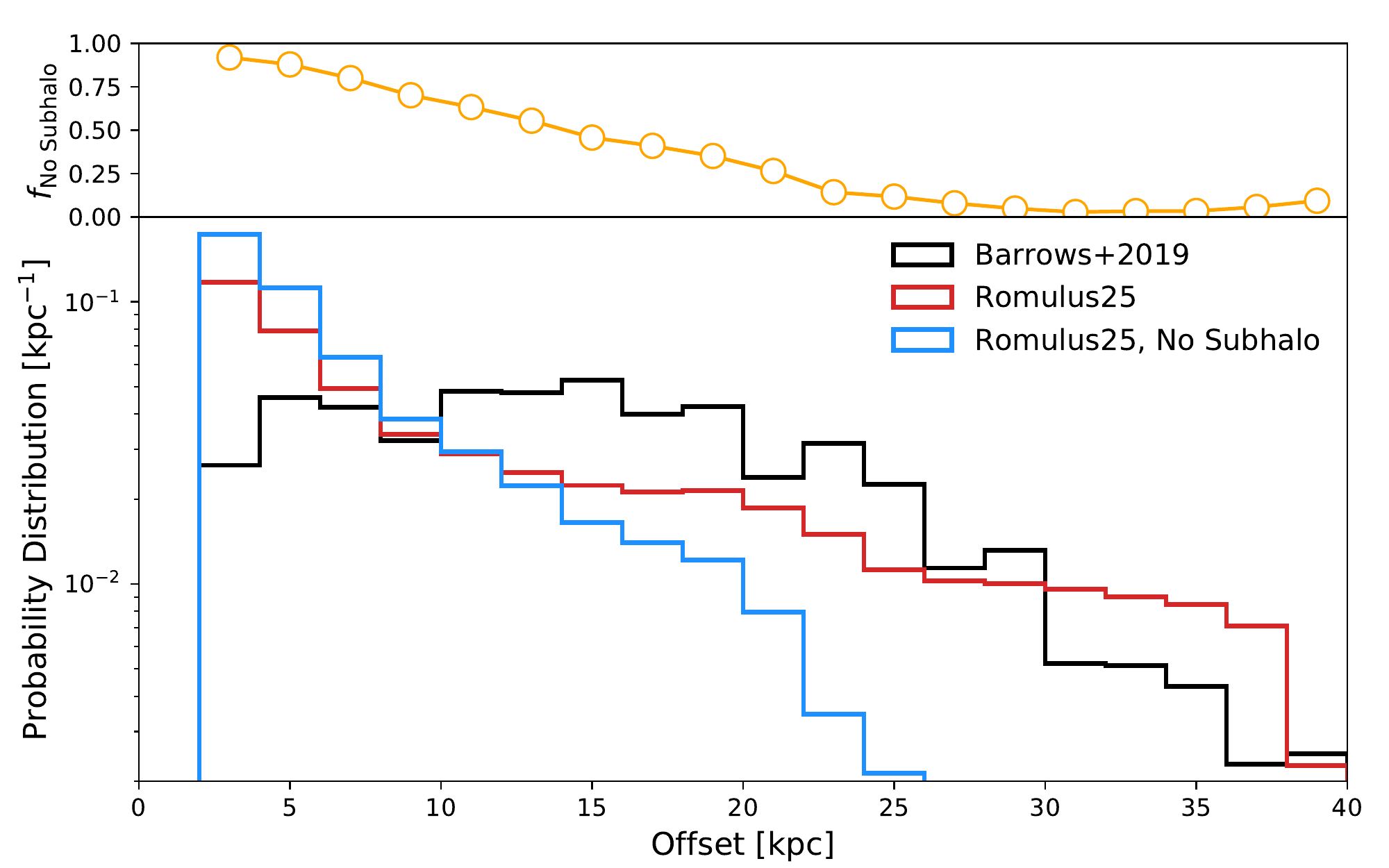}
  \caption{Comparison of {\sc Romulus25} predicted offset distributions to the detected HLX sample of \citet{Barrows+2019}, all normalized to integrate to unity.  In black, we plot the distribution of observed sources from \citet{Barrows+2019}.  In red, we plot the projected distribution of offsets from {\sc Romulus} emulating their selection, considering galaxies with $r<22.5$, $0.05<z<0.9$, and $L_\mathrm{bol}>10^{42} \ \mathrm{erg} \; \mathrm{s}^{-1}$ when averaged over 30 Myr.  We reproduce a broad distribution of offsets, but find that the true distribution should peak at small radii with sources likely missing due to limited astrometric precision.  In blue, we plot the distribution for the subset of wanderers that are not considered to reside within a resolved sub-halo (satellite galaxy) in the simulation, revealing that the most extreme offsets are most likely to arise from interloping faint galaxies.  In the upper panel, we plot the fraction of apparently offset SMBHs which lack an accompanying subhalo, and find that this function drops below 50\% at separations of $\gtrsim 15$ kpc. \label{fig:barrows_comparison}}
\end{figure*}

Accreting, wandering black holes, like those shown in Figure \ref{fig:example}, can manifest as Hyper-luminous X-ray sources (HLXs).  These are off-nuclear X-ray sources with X-ray luminosities above $10^{41} \ \mathrm{erg}\;\mathrm{s}^{-1}$ \citep{Matsumoto+2001,Kaaret+2001,Gao+2003}.  As the most extreme tail of the ultraluminous X-ray (ULX) population, these sources are unlikely to be produced by stellar mass objects and may instead represent accreting wandering SMBHs \citep{King&Dehnen2005}. Masses in the intermediate black hole mass range $(M_{\rm BH} < 10^6\,M_{\odot})$ are also supported by X-ray spectral fitting, which in turn imply relatively low blackbody temperatures \citep{Miller+2003,Davis+2011}.

\citet{Barrows+2019} assembled a sample of 169 HLXs by cross-matching galaxies with known redshifts in the Sloan Digital Sky Survey (SDSS) with the Chandra Source Catalog \citep{Evans+2010}. This sample includes sources with hard X-ray luminosities in excess of $10^{41} \ \mathrm{erg} \; \mathrm{s}^{-1}$ in the 2-10 keV band up to $z\sim 0.9$.  \citet{Barrows+2019} conservatively keep only those X-ray sources which are offset by $\geq 5$ times the uncertainty of their positions relative to their host galaxy centroids, which varies from source to source.  Most HLXs in this sample exhibit offsets of tens of kpc.

In Figure \ref{fig:barrows_comparison}, we compare the distribution of offsets obtained from this work with {\sc Romulus25}.  In black, we plot the distribution of offsets from \citet{Barrows+2019}, with each source weighted by the factor $(1-f)$, where $f$ is their estimated contamination fraction of each source. The red histogram is the distribution of offsets in the {\sc Romulus25} simulation obtained when emulating their selection criteria.  For this comparison, we first select galaxies with rest frame $r<22.5$ for all snapshots with redshifts $z<0.9$, to mimic the pre-selection of galaxies in the Sloan Digital Sky Survey (SDSS).  Then, we search for every SMBH within the virial radius of each selected halo, even if it is within a satellite halo.  If its bolometric luminosity averaged over the past 30 Myr is at least $10^{42} \ \mathrm{erg} \; \mathrm{s}^{-1}$, (that is, if its X-ray luminosity is at least $10^{41} \ \mathrm{erg} \; \mathrm{s}^{-1}$ assuming a bolometric correction of 10\%), we consider it detectable and save its offset from the center of the halo.  The three-dimensional spatial distribution of HLXs is then projected onto the sky using an Abel transform, and each redshift snapshot $n$ in the simulation between $0.05<z<0.9$ is volume-weighted by redshift.\footnote{Each redshift slice is assigned a weight $dV/dn = (dV/dz)(dz/dn)$, where $dV/dz=c d_L^2 (1+z)^{-1} dt/dz$ is the evolution of the comoving volume element with redshift, and $dz/dn$ is the frequency with which redshifts are sampled in the simulation outputs.  Here, $c$ is the speed of light and $d_L$ is the luminosity distance.}  These distributions are normalized to integrate to unity.

Comparing the red {\sc Romulus25} predicted distribution to the observed HLX distribution shown in black, we do indeed reproduce a population of luminous wandering SMBHs with separations of tens of kpc, as observed in \citet{Barrows+2019}. The most extreme offset SMBHs typically reside in the  outskirts of the halos of massive galaxies, with stellar masses $>10^{11.4} \ \mathrm{M}_\odot$.  However, unlike the observations, we find that this distribution should continue to increase with decreasing projected separation.  This implies the existence of a much larger population of moderately offset sources missed observationally by current campaigns \citep{Barrows+2019}.  This is most likely a selection effect due to limited astrometric precision, as discussed further in \citet{Stemo+2020}.  

Using the {\sc Amiga} halo finder, we can assess the fraction of detectable HLXs in the simulation that are ``true'' wanderers bereft of any resolved subhalo, as opposed to those which only appear as wanderers because they reside in nearby satellite galaxies too faint to have been detected.  In the blue histogram of Figure \ref{fig:barrows_comparison}, we repeat the analysis done to produce the red histogram, but omit any SMBHs which actually reside in a subhalo detected by our halo finder.  This distribution declines more rapidly with radius, indicating that the most offset HLXs likely reside in faint satellite galaxies.  We plot the fraction of wanderers lacking a subhalo in the simulation as a function of projected halo-centric distance in the upper panel.  We find that this fraction stays above 50\% until a projected separation of 15 kpc.  

\begin{figure*}
  \centering
  \includegraphics[width=\textwidth]{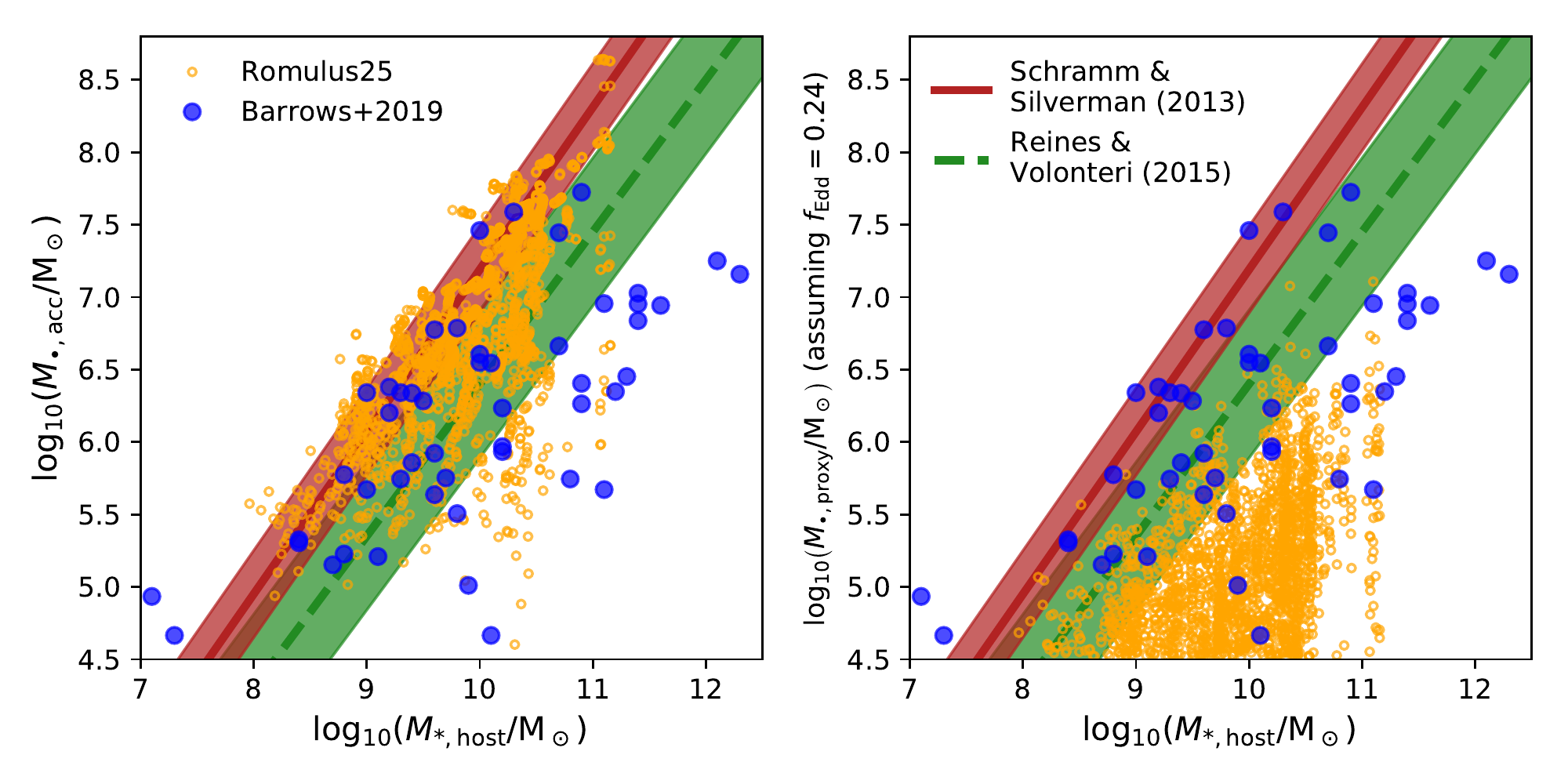}
  \caption{Comparison of the $M_\bullet-M_*$ relation for HLXs in the \citet{Barrows+2019} sample (blue) compared to {\sc Romulus25} (orange).  The \citet{Barrows+2019} sample plotted here only includes those HLXs with optically identifiable stellar counterparts in their study.  They found their sources to be broadly consistent with the relationship observed by \citet{Reines&Volonteri2015}, shown as a dashed green line.  In the left panel, we find that accreted masses of {\sc Romulus25} HLXs are also broadly consistent with typical galaxies, but they are calibrated to a different relationship, shown as the red solid line \citep{Schramm&Silverman2013}.  In \citet{Barrows+2019}, SMBH masses are estimated by assuming that they accrete with a uniform Eddington ratio of 0.24, motivated by their X-ray spectral indices.  In the right panel, we perform a more direct comparison with the {\sc Romulus} simulations by also applying this proxy and ``estimating'' SMBH masses based on their accretion rates, averaged over 30 Myr.  We find that this would have caused us to greatly underestimate {\sc Romulus25} SMBH masses, since {\sc Romulus} wanderers typically accrete at much lower Eddington ratio and are therefore fainter than these HLXs \citep{Ricarte+2019}. \label{fig:barrows_host_comparison}}
\end{figure*}

For HLXs with subsequently identified optical counterparts, \citet{Barrows+2019} also considered the SMBH mass ($M_\bullet$) to stellar mass ($M_*$) relation for their sources.  Lacking a more direct measurement of $M_\bullet$, they estimated $M_\bullet$ by assuming that the HLXs accrete with an Eddington ratio of 0.24, motivated by the average X-ray spectral index observed and a relationship between spectral index and Eddington ratio \citep{Greene&Ho2007}.  We perform the same selection on our {\sc Romulus25} HLXs which have an identifiable subhalo (those which are part of the inventory in the red histogram but not the blue one in Figure \ref{fig:barrows_comparison}).  In Figure \ref{fig:barrows_host_comparison}, we plot the SMBH mass to stellar mass relation for the subset of the \citet{Barrows+2019} sample with identified stellar counterparts as solid blue circles, where error-bars have been omitted for clarity.  We plot the SMBH mass as a function of host stellar mass in the {\sc Romulus25} simulation in orange, computed in a different way in each of the panels.  In the left panel, we plot each SMBH's accreted mass\footnote{A SMBH's accreted mass is the portion of mass acquired purely from gas accretion, excluding any number of seed masses that have merged to form the final product.} in the simulation, which has been found to trace the $M_\bullet-M_*$ relation in {\sc Romulus} even below the seed mass of $10^6 \ \mathrm{M}_\odot$ and can be interpreted as a proxy for the true black hole mass in dwarf galaxies \citep{Ricarte+2019}.  (See Appendix \ref{sec:raw_accreted_mass} and Figure \ref{fig:barrows_host_comparison_accreted_mass_check} for a comparison of accreted and raw SMBH masses among our wanderers.) \citet{Barrows+2019} found that their sources were broadly consistent with AGN in low-mass galaxies, the relationship found by \citep{Reines&Volonteri2015}, shown as a green dashed line.  Similarly, we find that the accreted masses of {\sc Romulus25} HLXs are also consistent with typical galaxies of the same mass.  Here, the appropriate comparison is actually the relationship found by \citet{Schramm&Silverman2013} shown in red, which is the relation used to calibrate the SMBH accretion and feedback parameters in {\sc Romulus}.  This reproduction is not entirely trivial, since environmental processes could have potentially modified this relationship for galaxies in these subhalos.

In the right panel, we emulate the \citet{Barrows+2019} proxy by ``estimating'' masses from SMBH luminosities assuming an Eddington ratio of 0.24.  We find that this would have caused us to greatly underestimate {\sc Romulus25} masses, since {\sc Romulus25} SMBHs tend to have much lower Eddington ratios than this adopted value \citep{Ricarte+2019}.  That is, {\sc Romulus} HLXs tend to be fainter than those in \citet{Barrows+2019} at a given SMBH mass.

The overall good agreement between Romulus predictions and the observed HLXs lends us confidence in the robustness of the simulated properties of the off-center SMBH population. In subsequent sections, we explore the key electromagnetic properties of the wandering population in Romulus. 

\subsection{Occurrence Rates of Dual AGN}
\label{sec:dual_agn}

\begin{figure*}
  \centering
  \includegraphics[width=\textwidth]{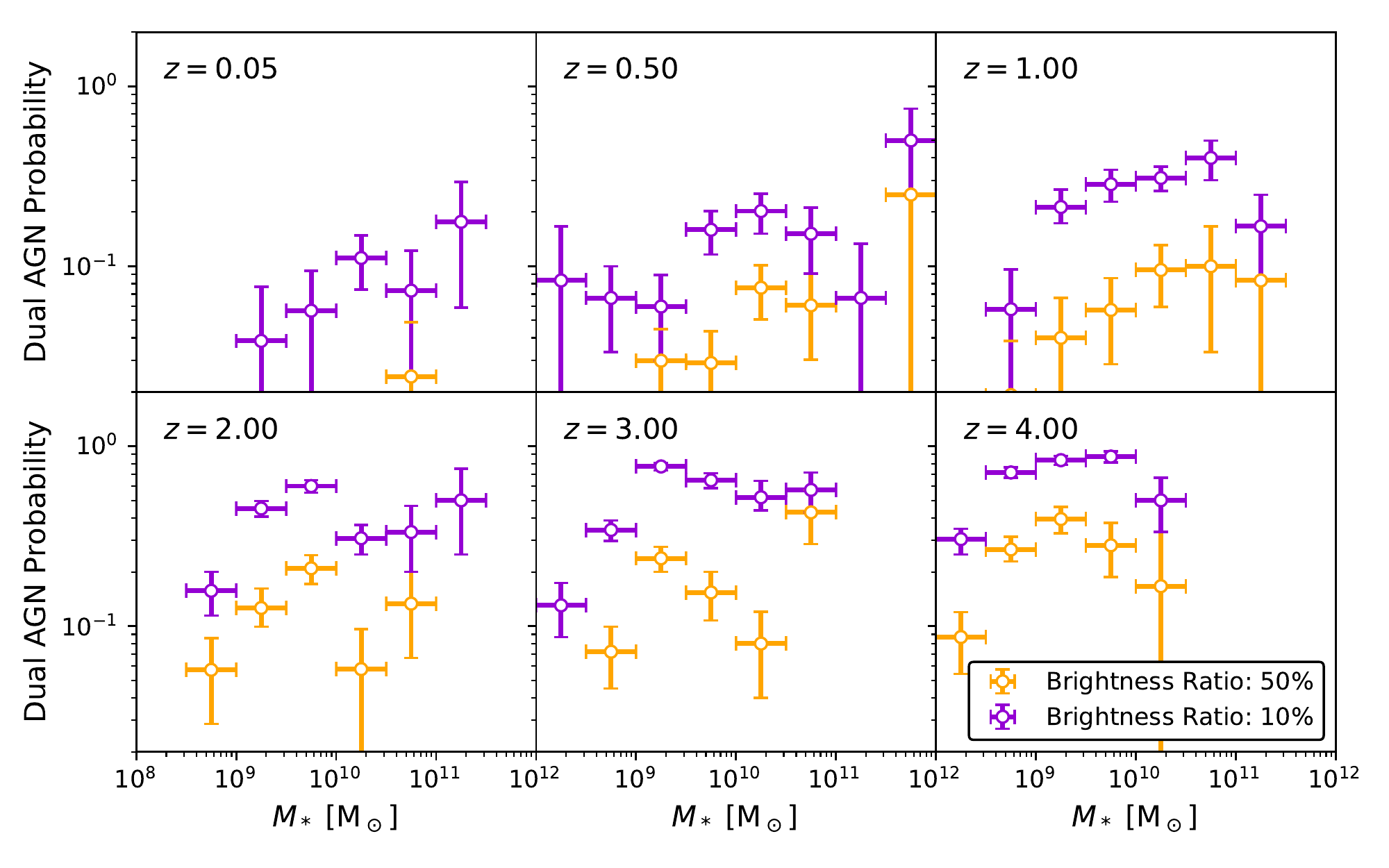}
  \caption{Given an AGN shining with a bolometric luminosity of $L_1 > 10^{42} \ \mathrm{erg} \; \mathrm{s}^{-1}$, we plot the probability of there existing a second AGN in the same halo shining with a luminosity $L_2 > R L_1$ as a function of stellar mass and redshift.  Two values of this luminosity ratio $R$ are considered, 50\% in orange and 10\% in purple.  We find that this probability increases with both stellar mass and redshift.  Errorbars originate from bootstrapping.  At $z=4$, {\it most} galaxies which host one SMBH above this threshold host another with at least 10\% of the first's luminosity, motivating deep surveys and sensitive instruments capable of testing this prediction.  \label{fig:luminosityRatios_42}}
\end{figure*}

Dual AGN are ubiquitous in the {\sc Romulus} universe, at least at low luminosities where we can build meaningful statistics.  In Figure \ref{fig:luminosityRatios_42}, we plot the probability of there being a second AGN in a galaxy with a (bolometric) luminosity at least $L_2 > RL_1$, given that there is already a first AGN with a luminosity of at least $L_1 > 10^{42} \ \mathrm{erg} \; \mathrm{s}^{-1}$.  We plot the luminosity of these wanderers as a function of host stellar mass across several redshifts.  Two brightness ratios are provided, $R=1/2$ in orange and $R=1/10$ in purple.  Error bars are estimated via bootstrapping in each stellar mass bin.  Although a bolometric luminosity threshold of $10^{42} \  \mathrm{erg} \; \mathrm{s}^{-1}$ is quite low, we are unfortunately limited by the relatively small (25 Mpc)$^3$ volume of {\sc Romulus25}, which also does not produce many high Eddington ratio systems \citep{Ricarte+2019}.

In {\sc Romulus25}, a galaxy hosting an SMBH with $L_\mathrm{bol}>10^{42} \ \mathrm{erg}\;\mathrm{s}^{-1}$ at $z=0.05$ has a roughly 10\% chance of hosting a second SMBH shining with at least a tenth the first's luminosity.  However, we predict that dual AGN should be much more common at higher redshifts, motivating deep surveys and sensitive instruments capable of testing this prediction.  At present, searches for spatially-resolved dual AGN in the X-ray are confined to $z\lesssim 0.05$ and $R\gtrsim 0.01$ \citep{Foord+2019,Foord+2021}.  Yet at $z=4$, the {\it majority} of galaxies in {\sc Romulus} with one SMBH above this luminosity threshold host another with at least 10\% of its luminosity.  We have tested increasing the luminosity threshold to $L_1 = 10^{43} \ \mathrm{erg}\;\mathrm{s}^{-1}$, and did not find any noticeable difference aside from poorer statistics.  SMBH accretion rates are averaged over 30 Myr in Figure \ref{fig:luminosityRatios_42}, but it is possible for AGN to have shorter duty cycles than this.  To interpret these results assuming that the AGN only shine a fraction $D$ of the time, $L_1$ should be multiplied by a factor $1/D$ to conserve the average accretion rate, but the dual AGN probability must be correspondingly divided by the same factor.  

\subsection{Average Wanderer Emission Profiles}
\label{sec:profile}

\begin{figure*}
  \centering
  \includegraphics[width=\textwidth]{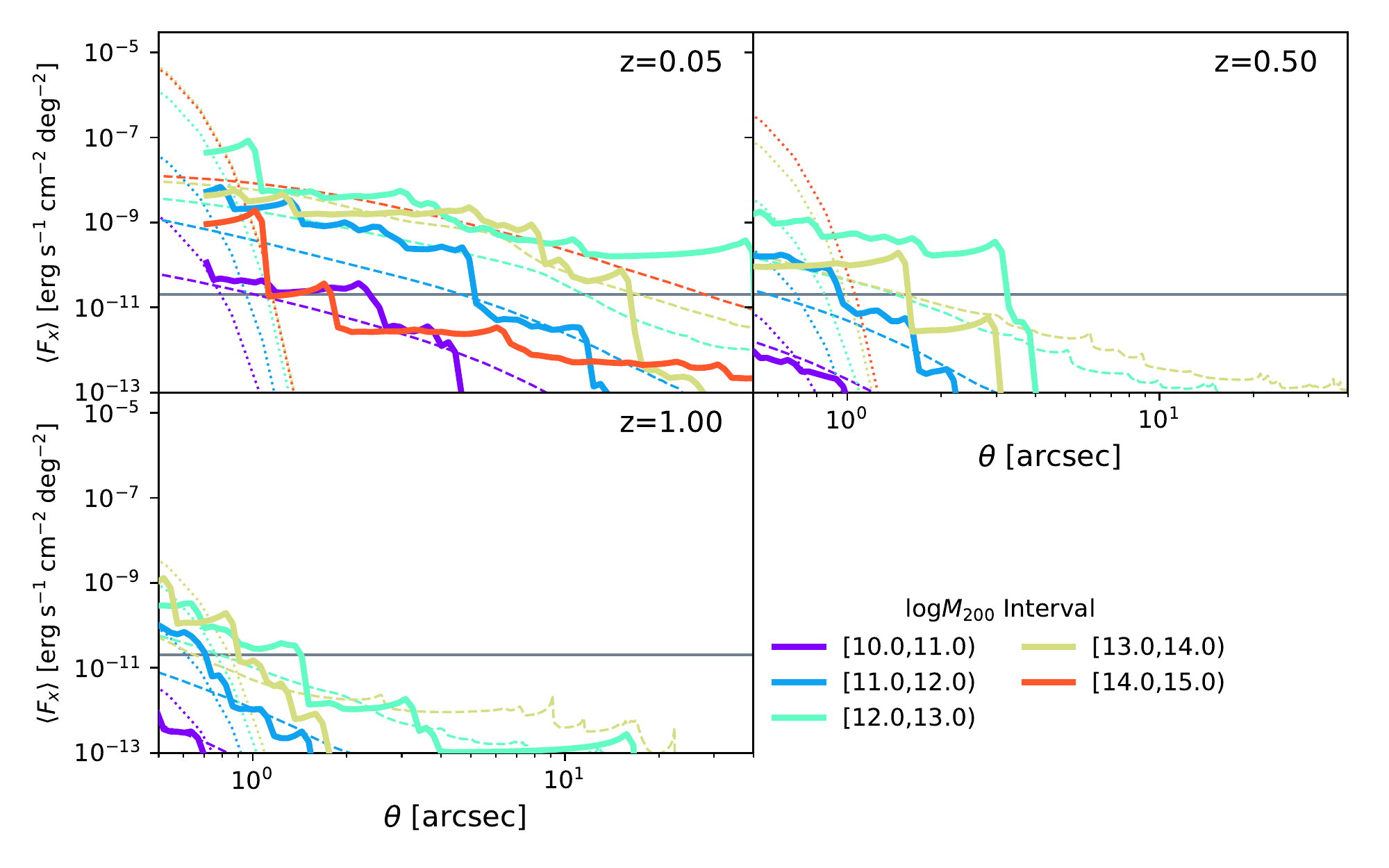}
  \caption{Predicted X-ray emission profiles due to accreting wandering SMBHs for different halo masses and redshifts. Solid colored lines plot 2-10 keV emission profiles from wandering SMBHs calculated by assuming a bolometric correction of 0.1 and averaging halos of similar mass.  With thin dotted lines, we plot the average flux of central SMBHs among the same halos, convolved with a two-dimensional Gaussian point spread function with a full width at half maximum of 0.5 arcseconds, appropriate for {\it Chandra}.  Shown as thin dashed lines, we also estimate the contribution from XRBs based on empirical relations.  Colors correspond to the halo masses averaged in each curve, as indicated in the legend.  A grey horizontal line marks the cosmic X-ray background level in this X-ray band \citep{Cappelluti+2017}.  At small redshifts, {\sc Romulus} predicts a profile of X-ray emission due to wandering SMBHs that is most strongly above both XRBs and the X-ray background in the halo mass range between $10^{11-13} \ \mathrm{M}_\odot$. 
  \label{fig:emissionProfile}}
\end{figure*}

If wandering SMBHs prove to be difficult to find and resolve individually, they may likely manifest as a ``halo'' of emission or excess counts in stacked images.  In Figure \ref{fig:emissionProfile}, we compute the average emission profile of wandering SMBHs that one would obtain by stacking halos of similar mass.  Once again, we average wandering SMBH accretion rates over a 30 Myr timescale, and further assume that 10\% of the radiated energy is emitted in the 2-10 keV X-ray band \citep[e.g.,][]{Hopkins+2007}.  We perform an Abel transform to project average spherical profiles onto the plane of the sky, and apply the appropriate cosmological factors due to the evolving luminosity and angular diameter distances with redshift to derive this estimate.  Each color represents a different bin in halo mass, as indicated by the legend.

At $z=0.05$, halos in the $10^{11-13} \ \mathrm{M}_\odot$ mass range exhibit average wandering light profiles orders of magnitude above the cosmic X-ray background, plotted in gray.  The solid horizontal line represents the cosmic X-ray background levels in the 2.0-10.0 keV band \citep{Cappelluti+2017}.  Dwarf galaxy halos (purple) have too few wanderers with too little luminosity, while the galaxy cluster halo in {\sc RomulusC} (orange) exhibits suppressed emission, since most galaxies are quenched and quiescent in the cluster environment \citep{Tremmel+2019}.  Unfortunately, although the {\sc RomulusC} cluster hosts 1613 wandering SMBHs, they accrete less efficiently in its hotter halo \citep[see also][for further discussion of this phenomenon]{Ricarte+2021}.  

As thin dotted curves, we plot the average flux from central SMBHs, blurred assuming a Gaussian point-spread function with a full width at half maximum of 0.5 arcseconds, appropriate for the center of the {\it Chandra} field of view.  Additionally, as thin dashed curves, we plot the expected contribution from X-ray binaries (XRBs), for which details are provided in Appendix \ref{sec:hlxs}.  At low enough redshift, these faint, extended halos of wanderer emission are much larger than the {\it Chandra} PSF and may extend to $\sim$10 arcseconds above both the X-ray background and host galaxies' XRBs.  However, wanderers become too faint to detect above X-ray background levels by $z=1$.  This signal could potentially be detected by stacking X-ray observations centered on similar mass galaxies, masking out satellites, and comparing the resulting profiles to expectations from the field.

\subsection{Wandering Tidal Disruption Events?}
\label{sec:tdes}

\begin{figure*}
  \centering
  \includegraphics[width=\textwidth]{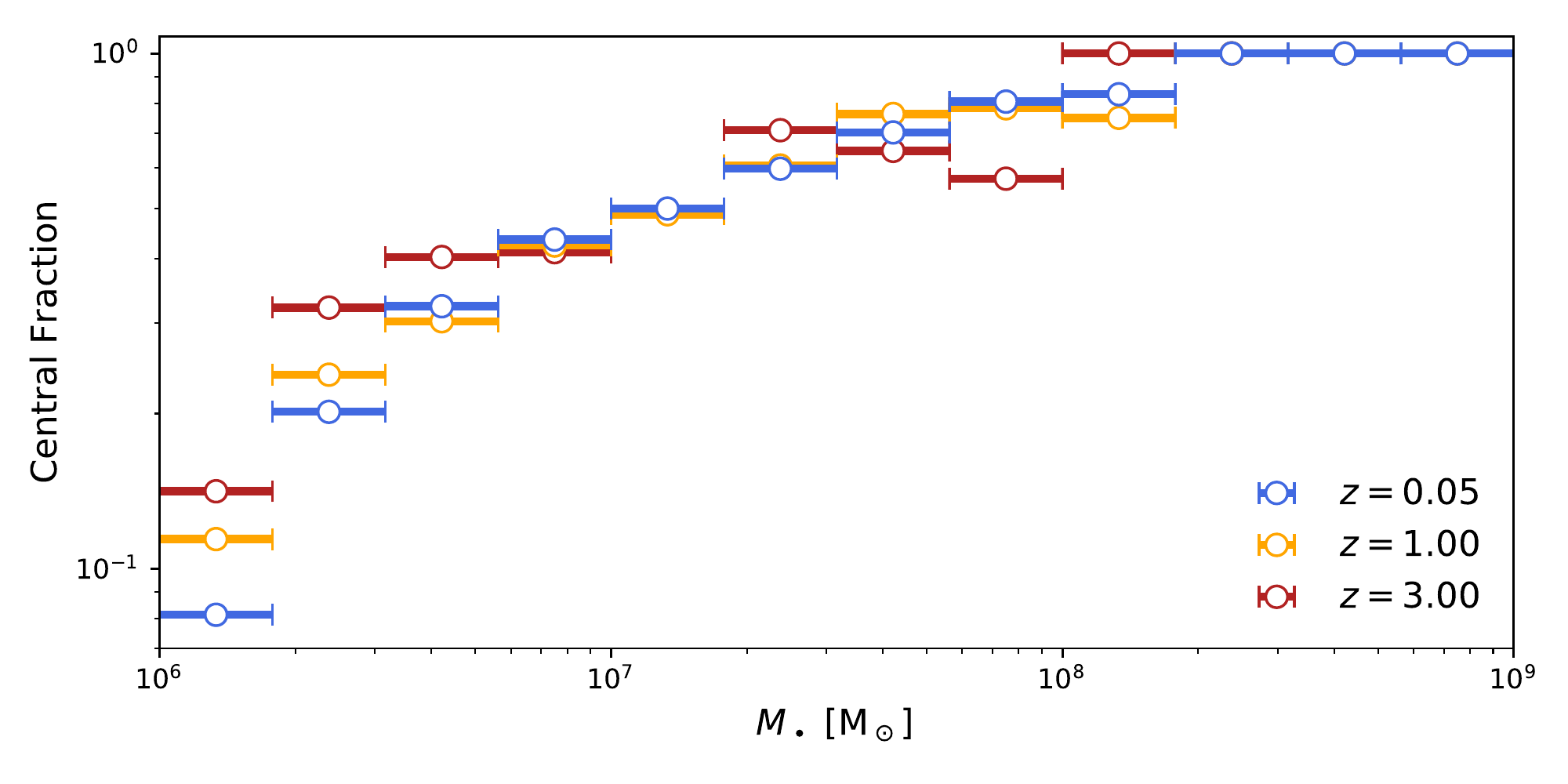}
  \caption{The fraction of SMBHs of a given mass which are centrally located in {\sc Romulus25}.  This fraction drops far below unity at low masses, meaning that the typical $10^6 \ \mathrm{M}_\odot$ SMBH is actually a wanderer.  If these wandering SMBHs are able to retain an unresolved stellar component, they may dominate the TDE rate in their mass bin.  \label{fig:tde_correction}}
\end{figure*}

A TDE flare is caused by the destruction and accretion of a star torn apart by tidal forces when it approaches too close to a SMBH \citep{Rees1988}.  Several tens of TDEs have been observationally identified, and soon orders of magnitude more are expected from ongoing and future surveys such as the Zwicky Transient Factory and Vera Rubin Observatory \citep{Strubbe&Quataert2009,vanVelzen+2011,Bricman&Gomboc2020}.  The number density of SMBHs as a function of mass is the basic ingredient needed to compute theoretical estimates of the volumetric TDE rate, and this has typically been derived from scaling relations between {\it central} SMBH masses and host galaxy properties \citep{Stone&Metzger2016,Stone+2020}.  In this section, we demonstrate that the majority of SMBHs with masses $\lesssim 10^7 \ \mathrm{M}_\odot$ are in fact wanderers, missed entirely by this kind of accounting.  If these wandering SMBHs can retain stellar counterparts (such as their nuclear star clusters, NSCs) and disrupt stars at comparable rates, then the majority of TDEs originating from SMBHs with $M_\bullet < 10^7 \ \mathrm{M}_\odot$ may be centrally offset.  

In Figure \ref{fig:tde_correction}, we plot the fraction of SMBHs of a given mass in {\sc Romulus25} that are centrals, for three different redshifts.  We find that this fraction drops rapidly at low masses, such that at $z=0.05$, only one in twelve $10^6 \ \mathrm{M}_\odot$ SMBHs are centrally located.  The central fraction decreases with decreasing redshift, as more minor mergers build up the wandering population.  If (i) these statistics are representative of the real Universe, (ii) wandering SMBHs disrupt stars at a comparable rate to centrals, and (iii) offset TDEs are equally identifiable, then the majority of TDEs due to SMBHs with $M_\bullet < 10^7 \ \mathrm{M}_\odot$ should be offset from their galactic centers.  

This may require a significant upwards revision of the theoretically estimated TDE rates, which we estimate here. \citet{Stone&Metzger2016} express the theoretical TDE rate as a function of SMBH mass as $N(M_\bullet/10^8 \ \mathrm{M}_\odot)^B$, where $N=2.9\times 10^{-5} \ \mathrm{yr}^{-1}$ and $B=-0.404$.  Consequently, we can estimate this correction factor via $C=\sum_\mathrm{all} M_\bullet^B / \sum_\mathrm{cen} M_\bullet^B$, representing a sum over all SMBHs in {\sc Romulus25} in the numerator and a sum restricted to central SMBHs in the denominator.  We find that the volumetric TDE rate when accounting for wandering SMBHs should be revised upwards by a factor of $C=8.0$ at $z=0.05$, owing mostly to $10^6 \ \mathrm{M}_\odot$ SMBHs, to which current surveys are not sensitive.  This may even be an underestimate, since the black hole occupation fraction in dwarf galaxies is highly uncertain and may well be unity \citep[e.g.,][]{Mezcua+2018,Baldassare+2020}.  At present, there exists one compelling offset TDE candidate, which is located 12.5 kpc from the center of a lenticular galaxy and exhibits a $L \propto t^{-5/3}$ light curve characteristic of TDEs \citep{Lin+2018}.  The occurrence rate of offset TDEs can place joint constraints on SMBH dynamics, the SMBH occupation fraction, and the NSC occupation fraction.

\section{Discussion}

We present observational comparisons and predictions for wandering SMBHs using the {\sc Romulus} suite of cosmological simulations. These simulations carefully apply a corrective dynamical friction force onto SMBHs to produce realistic dynamics and implement a physically motivated seeding prescription that produces SMBH occupation fractions consistent with observations. Our key findings are summarized as follows:

\begin{itemize}
    \item We compare the wandering population of {\sc Romulus25} to the \citet{Barrows+2019} sample of hyperluminous X-ray sources (HLXs).  We reproduce a broad offset distribution, and find that most objects offset by more than 15 kpc are likely to be in satellite galaxies.  The population of greatly offset HLXs in this sample also implies a much larger population with more modest offsets, which are more likely to lack a corresponding subhalo.
    \item Dual AGN are common in the {\sc Romulus} universe at the low luminosities that these simulations can probe.  The majority of galaxies at $z=4$ containing an AGN shining above $L_\mathrm{bol}>10^{42} \ \mathrm{erg}\;\mathrm{s}^{-1}$ also contain a second AGN of comparable brightness.
    \item Wandering SMBHs may collectively manifest as an overdensity of X-ray emission around galaxies in excess of the cosmic background.  Stacked X-ray observations of galaxies may reveal a faint halo of emission attributable to these wanderers.
    \item Below $10^7 \ \mathrm{M}_\odot$, central SMBHs of a given mass are greatly outnumbered by wanderers.  If these wandering SMBHs can retain their nuclear star clusters, wanderers may dominate the TDE rate by low-mass SMBHs.
\end{itemize}

An important caveat for this work is the underlying assumption of Bondi-like accretion, a sub-grid prescription used in most cosmological simulations.  Higher resolution simulations have demonstrated that the accretion rate onto wanderers may be limited to $\sim 10-20\%$ the Bondi rate due to the wide distribution of angular momentum encountered moving through the halo \citep{Guo+2020}.  Furthermore, if SMBH accretion rates are low enough for their disks to be in the advection dominated regime, then the accumulation of magnetic flux around their horizons may further suppress accretion rates by orders of magnitude \citep{Igumenshchev&Narayan2002,Perna+2003,Pellegrini2005,Ressler+2021}.  The accretion and feedback prescriptions in these simulations successfully reproduce the empirical $M_\bullet-M_*$ scaling relation \citep{Schramm&Silverman2013}, and we have previously shown that the typical luminous wanderer may have an Eddington ratio large enough to avoid forming an advection domination accretion flow \citep{Ricarte+2021}.  Nevertheless, as numerical techniques improve, it will be useful to compare the results from {\sc Romulus} to other cosmological simulations with different sub-grid physics.

These simulations predict that dual AGN may be common at low luminosities and high redshifts, motivating deep studies that can test this prediction.  Additional detailed comparisons of offset/dual AGN samples that carefully forward-model selection effects would be useful to calibrate theoretical uncertainties, such as the temporally unresolved AGN duty cycle.  In our current work, we have not considered the velocities of wandering SMBHs. In galactic nuclei, velocities of the measured double-peaked emission lines have been used to spectrally identify dual/offset AGN \citep[e.g.,][]{Blecha+2013,Comerford&Greene2014,Pesce+2021}.

\citet{Ricarte+2021} found that most wandering SMBHs in {\sc Romulus} do not reside in {\it resolved} stellar overdensities, except for those at large radii which experience weaker tidal forces.  However, our simulations cannot resolve the formation and disruption of structures below the gravitational softening length of 350 pc.  Below these scales, SMBHs are expected to be accompanied by NSCs, especially in the mass ranges spanned by wandering SMBHs \citep{Neumayer+2020}. \citet{vandenBosch+2018} argue that it is extremely difficult to completely disrupt the central regions of a halo in the cold dark matter paradigm.

Due to limited resolution, our simulations lack two important channels to create wanderers, namely multi-body SMBH interactions \citep[e.g.,][]{Volonteri+2003} and gravitational wave recoil following SMBH mergers \citep[e.g.,][]{Libeskind+2006,Volonteri2007,Holley-Bockelmann+2008}.  These types of wanderers may dominate the wandering population at low halo-centric radius \citep{Volonteri&Perna2005,Izquierdo-Villalba+2020}.

In conclusion, simulations predict the existence of an extensive wandering SMBH population that stands to be revealed via their electromagnetic signatures. This work suggests that our current census of detected SMBHs is highly incomplete. 

\section{Acknowledgments}

We thank our anonymous referee for insightful comments which improved the presentation of our results. AR thanks Ramesh Narayan for continued support and insightful conversations about wandering SMBH accretion rates.  We thank Nicholas Stone for fruitful conversations about TDEs, and Kristen Garofali for helpful conversations about X-ray emission of hot gas.

AR is supported by the National Science Foundation under Grant No. OISE 1743747 as well as the Black Hole Initiative (BHI) by grants from the Gordon and Betty Moore Foundation and the John Templeton Foundation. MT is supported by an NSF Astronomy and Astrophysics Postdoctoral Fellowship under award AST-2001810. PN acknowledges hospitality and support from the Aspen Center for Physics, where some of the early ideas that informed this work were developed during the summer workshop titled "Emergence, Evolution and Effects of Black Holes in the Universe: The Next 50 Years of Black Hole Physics" that she co-organized. The Aspen Center for Physics, is supported by National Science Foundation grant PHY-1066293. The {\sc Romulus} simulations are part of the Blue Waters sustained-petascale computing project, which is supported by the National Science Foundation (awards OCI-0725070 and ACI-1238993) and the state of Illinois. Blue Waters is a joint effort of the University of Illinois at Urbana–Champaign and its National Center for Supercomputing Applications.  This work is also part of a Petascale Computing Resource Allocations allocation support by the National Science Foundation (award number OAC-1613674). This work also used the Extreme Science and Engineering Discovery Environment (XSEDE), which is supported by National Science Foundation grant number ACI-1548562. Resources supporting this work were also provided by the NASA High-End Computing (HEC) Program through the NASA Advanced Supercomputing (NAS) Division at Ames Research Center.  Analysis was conducted on the NASA Pleiades computer and facilities supported by the Yale Center for Research Computing.

\section{Data Availability}

The data presented in this article will be available upon request.

\bibliography{ms}

\begin{thebibliography}{}
\expandafter\ifx\csname natexlab\endcsname\relax\def\natexlab#1{#1}\fi
\providecommand{\url}[1]{\href{#1}{#1}}
\providecommand{\dodoi}[1]{doi:~\href{http://doi.org/#1}{\nolinkurl{#1}}}
\providecommand{\doeprint}[1]{\href{http://ascl.net/#1}{\nolinkurl{http://ascl.net/#1}}}
\providecommand{\doarXiv}[1]{\href{https://arxiv.org/abs/#1}{\nolinkurl{https://arxiv.org/abs/#1}}}

\bibitem[{{Alexander} \& {Natarajan}(2014)}]{Alexander&Natarajan2014}
{Alexander}, T., \& {Natarajan}, P. 2014, Science, 345, 1330,
  \dodoi{10.1126/science.1251053}

\bibitem[{{Amaro-Seoane} {et~al.}(2017){Amaro-Seoane}, {Audley}, {Babak},
  {Baker}, {Barausse}, {Bender}, {Berti}, {Binetruy}, {Born}, {Bortoluzzi},
  {Camp}, {Caprini}, {Cardoso}, {Colpi}, {Conklin}, {Cornish}, {Cutler},
  {Danzmann}, {Dolesi}, {Ferraioli}, {Ferroni}, {Fitzsimons}, {Gair}, {Gesa
  Bote}, {Giardini}, {Gibert}, {Grimani}, {Halloin}, {Heinzel}, {Hertog},
  {Hewitson}, {Holley-Bockelmann}, {Hollington}, {Hueller}, {Inchauspe},
  {Jetzer}, {Karnesis}, {Killow}, {Klein}, {Klipstein}, {Korsakova}, {Larson},
  {Livas}, {Lloro}, {Man}, {Mance}, {Martino}, {Mateos}, {McKenzie},
  {McWilliams}, {Miller}, {Mueller}, {Nardini}, {Nelemans}, {Nofrarias},
  {Petiteau}, {Pivato}, {Plagnol}, {Porter}, {Reiche}, {Robertson},
  {Robertson}, {Rossi}, {Russano}, {Schutz}, {Sesana}, {Shoemaker}, {Slutsky},
  {Sopuerta}, {Sumner}, {Tamanini}, {Thorpe}, {Troebs}, {Vallisneri},
  {Vecchio}, {Vetrugno}, {Vitale}, {Volonteri}, {Wanner}, {Ward}, {Wass},
  {Weber}, {Ziemer}, \& {Zweifel}}]{LISA+2017}
{Amaro-Seoane}, P., {Audley}, H., {Babak}, S., {et~al.} 2017, arXiv e-prints,
  arXiv:1702.00786.
\newblock \doarXiv{1702.00786}

\bibitem[{{Armitage} \& {Natarajan}(2002)}]{Armitage+PN2002}
{Armitage}, P.~J., \& {Natarajan}, P. 2002, \apjl, 567, L9,
  \dodoi{10.1086/339770}

\bibitem[{{Arzoumanian} {et~al.}(2020){Arzoumanian}, {Baker}, {Blumer},
  {B{\'e}csy}, {Brazier}, {Brook}, {Burke-Spolaor}, {Chatterjee}, {Chen},
  {Cordes}, {Cornish}, {Crawford}, {Cromartie}, {Decesar}, {Demorest}, {Dolch},
  {Ellis}, {Ferrara}, {Fiore}, {Fonseca}, {Garver-Daniels}, {Gentile}, {Good},
  {Hazboun}, {Holgado}, {Islo}, {Jennings}, {Jones}, {Kaiser}, {Kaplan},
  {Kelley}, {Key}, {Laal}, {Lam}, {Lazio}, {Lorimer}, {Luo}, {Lynch},
  {Madison}, {McLaughlin}, {Mingarelli}, {Ng}, {Nice}, {Pennucci}, {Pol},
  {Ransom}, {Ray}, {Shapiro-Albert}, {Siemens}, {Simon}, {Spiewak}, {Stairs},
  {Stinebring}, {Stovall}, {Sun}, {Swiggum}, {Taylor}, {Turner}, {Vallisneri},
  {Vigeland}, {Witt}, \& {Nanograv Collaboration}}]{NANOGrav2020}
{Arzoumanian}, Z., {Baker}, P.~T., {Blumer}, H., {et~al.} 2020, \apjl, 905,
  L34, \dodoi{10.3847/2041-8213/abd401}

\bibitem[{{Baldassare} {et~al.}(2020){Baldassare}, {Geha}, \&
  {Greene}}]{Baldassare+2020}
{Baldassare}, V.~F., {Geha}, M., \& {Greene}, J. 2020, \apj, 896, 10,
  \dodoi{10.3847/1538-4357/ab8936}

\bibitem[{{Banik} {et~al.}(2019){Banik}, {van den Bosch}, {Tremmel}, {More},
  {Despali}, {More}, {Vegetti}, \& {McKean}}]{Banik+2019}
{Banik}, U., {van den Bosch}, F.~C., {Tremmel}, M., {et~al.} 2019, \mnras, 483,
  1558, \dodoi{10.1093/mnras/sty3267}

\bibitem[{{Barausse} {et~al.}(2020){Barausse}, {Dvorkin}, {Tremmel},
  {Volonteri}, \& {Bonetti}}]{Barausse+2020}
{Barausse}, E., {Dvorkin}, I., {Tremmel}, M., {Volonteri}, M., \& {Bonetti}, M.
  2020, \apj, 904, 16, \dodoi{10.3847/1538-4357/abba7f}

\bibitem[{{Barrows} {et~al.}(2019){Barrows}, {Mezcua}, \&
  {Comerford}}]{Barrows+2019}
{Barrows}, R.~S., {Mezcua}, M., \& {Comerford}, J.~M. 2019, \apj, 882, 181,
  \dodoi{10.3847/1538-4357/ab338a}

\bibitem[{{Bartlett} {et~al.}(2021){Bartlett}, {Desmond}, {Devriendt},
  {Ferreira}, \& {Slyz}}]{Bartlett+2021}
{Bartlett}, D.~J., {Desmond}, H., {Devriendt}, J., {Ferreira}, P.~G., \&
  {Slyz}, A. 2021, \mnras, 500, 4639, \dodoi{10.1093/mnras/staa3516}

\bibitem[{{Begelman} {et~al.}(1980){Begelman}, {Blandford}, \&
  {Rees}}]{Begelman+1980}
{Begelman}, M.~C., {Blandford}, R.~D., \& {Rees}, M.~J. 1980, \nat, 287, 307,
  \dodoi{10.1038/287307a0}

\bibitem[{{Bellovary} {et~al.}(2010){Bellovary}, {Governato}, {Quinn},
  {Wadsley}, {Shen}, \& {Volonteri}}]{Bellovary+2010}
{Bellovary}, J.~M., {Governato}, F., {Quinn}, T.~R., {et~al.} 2010, \apjl, 721,
  L148, \dodoi{10.1088/2041-8205/721/2/L148}

\bibitem[{{Bellovary} {et~al.}(2021){Bellovary}, {Hayoune}, {Chafla},
  {Vincent}, {Brooks}, {Christensen}, {Munshi}, {Tremmel}, {Quinn}, {Van Nest},
  {Sligh}, \& {Luzuriaga}}]{Bellovary+2021}
{Bellovary}, J.~M., {Hayoune}, S., {Chafla}, K., {et~al.} 2021, arXiv e-prints,
  arXiv:2102.09566.
\newblock \doarXiv{2102.09566}

\bibitem[{{Biernacki} {et~al.}(2017){Biernacki}, {Teyssier}, \&
  {Bleuler}}]{Biernacki+2017}
{Biernacki}, P., {Teyssier}, R., \& {Bleuler}, A. 2017, \mnras, 469, 295,
  \dodoi{10.1093/mnras/stx845}

\bibitem[{{Blecha} {et~al.}(2013){Blecha}, {Loeb}, \& {Narayan}}]{Blecha+2013}
{Blecha}, L., {Loeb}, A., \& {Narayan}, R. 2013, \mnras, 429, 2594,
  \dodoi{10.1093/mnras/sts533}

\bibitem[{{Bondi}(1952)}]{Bondi1952}
{Bondi}, H. 1952, \mnras, 112, 195, \dodoi{10.1093/mnras/112.2.195}

\bibitem[{{Booth} \& {Schaye}(2009)}]{Booth+Schaye2009}
{Booth}, C.~M., \& {Schaye}, J. 2009, \mnras, 398, 53,
  \dodoi{10.1111/j.1365-2966.2009.15043.x}

\bibitem[{{Bortolas} {et~al.}(2020){Bortolas}, {Capelo}, {Zana}, {Mayer},
  {Bonetti}, {Dotti}, {Davies}, \& {Madau}}]{Bortolas+2020}
{Bortolas}, E., {Capelo}, P.~R., {Zana}, T., {et~al.} 2020, \mnras, 498, 3601,
  \dodoi{10.1093/mnras/staa2628}

\bibitem[{{Bricman} \& {Gomboc}(2020)}]{Bricman&Gomboc2020}
{Bricman}, K., \& {Gomboc}, A. 2020, \apj, 890, 73,
  \dodoi{10.3847/1538-4357/ab6989}

\bibitem[{{Butsky} {et~al.}(2019){Butsky}, {Burchett}, {Nagai}, {Tremmel},
  {Quinn}, \& {Werk}}]{Butsky+2019}
{Butsky}, I.~S., {Burchett}, J.~N., {Nagai}, D., {et~al.} 2019, \mnras, 490,
  4292, \dodoi{10.1093/mnras/stz2859}

\bibitem[{{Cappelluti} {et~al.}(2017){Cappelluti}, {Li}, {Ricarte}, {Agarwal},
  {Allevato}, {Tasnim Ananna}, {Ajello}, {Civano}, {Comastri}, {Elvis},
  {Finoguenov}, {Gilli}, {Hasinger}, {Marchesi}, {Natarajan}, {Pacucci},
  {Treister}, \& {Urry}}]{Cappelluti+2017}
{Cappelluti}, N., {Li}, Y., {Ricarte}, A., {et~al.} 2017, \apj, 837, 19,
  \dodoi{10.3847/1538-4357/aa5ea4}

\bibitem[{{Chadayammuri} {et~al.}(2020){Chadayammuri}, {Tremmel}, {Nagai},
  {Babul}, \& {Quinn}}]{Chadayammuri+2020}
{Chadayammuri}, U., {Tremmel}, M., {Nagai}, D., {Babul}, A., \& {Quinn}, T.
  2020, arXiv e-prints, arXiv:2001.06532.
\newblock \doarXiv{2001.06532}

\bibitem[{{Chandrasekhar}(1943)}]{Chandrasekhar1943}
{Chandrasekhar}, S. 1943, \apj, 97, 255, \dodoi{10.1086/144517}

\bibitem[{{Chen} {et~al.}(2021){Chen}, {Ni}, {Tremmel}, {Di Matteo}, {Bird},
  {DeGraf}, \& {Feng}}]{Chen+2021}
{Chen}, N., {Ni}, Y., {Tremmel}, M., {et~al.} 2021, arXiv e-prints,
  arXiv:2104.00021.
\newblock \doarXiv{2104.00021}

\bibitem[{{Comerford} \& {Greene}(2014)}]{Comerford&Greene2014}
{Comerford}, J.~M., \& {Greene}, J.~E. 2014, \apj, 789, 112,
  \dodoi{10.1088/0004-637X/789/2/112}

\bibitem[{{Comerford} {et~al.}(2009){Comerford}, {Gerke}, {Newman}, {Davis},
  {Yan}, {Cooper}, {Faber}, {Koo}, {Coil}, {Rosario}, \&
  {Dutton}}]{Comerford+2009}
{Comerford}, J.~M., {Gerke}, B.~F., {Newman}, J.~A., {et~al.} 2009, \apj, 698,
  956, \dodoi{10.1088/0004-637X/698/1/956}

\bibitem[{{Davis} {et~al.}(2011){Davis}, {Narayan}, {Zhu}, {Barret}, {Farrell},
  {Godet}, {Servillat}, \& {Webb}}]{Davis+2011}
{Davis}, S.~W., {Narayan}, R., {Zhu}, Y., {et~al.} 2011, \apj, 734, 111,
  \dodoi{10.1088/0004-637X/734/2/111}

\bibitem[{{Dubois} {et~al.}(2014){Dubois}, {Pichon}, {Welker}, {Le Borgne},
  {Devriendt}, {Laigle}, {Codis}, {Pogosyan}, {Arnouts}, {Benabed}, {Bertin},
  {Blaizot}, {Bouchet}, {Cardoso}, {Colombi}, {de Lapparent}, {Desjacques},
  {Gavazzi}, {Kassin}, {Kimm}, {McCracken}, {Milliard}, {Peirani}, {Prunet},
  {Rouberol}, {Silk}, {Slyz}, {Sousbie}, {Teyssier}, {Tresse}, {Treyer},
  {Vibert}, \& {Volonteri}}]{Dubois+2014}
{Dubois}, Y., {Pichon}, C., {Welker}, C., {et~al.} 2014, \mnras, 444, 1453,
  \dodoi{10.1093/mnras/stu1227}

\bibitem[{{Evans} {et~al.}(2010){Evans}, {Primini}, {Glotfelty}, {Anderson},
  {Bonaventura}, {Chen}, {Davis}, {Doe}, {Evans}, {Fabbiano}, {Galle}, {Gibbs},
  {Grier}, {Hain}, {Hall}, {Harbo}, {He}, {Houck}, {Karovska}, {Kashyap},
  {Lauer}, {McCollough}, {McDowell}, {Miller}, {Mitschang}, {Morgan},
  {Mossman}, {Nichols}, {Nowak}, {Plummer}, {Refsdal}, {Rots}, {Siemiginowska},
  {Sundheim}, {Tibbetts}, {Van Stone}, {Winkelman}, \& {Zografou}}]{Evans+2010}
{Evans}, I.~N., {Primini}, F.~A., {Glotfelty}, K.~J., {et~al.} 2010, \apjs,
  189, 37, \dodoi{10.1088/0067-0049/189/1/37}

\bibitem[{{Foord} {et~al.}(2021){Foord}, {G{\"u}ltekin}, {Runnoe}, \&
  {Koss}}]{Foord+2021}
{Foord}, A., {G{\"u}ltekin}, K., {Runnoe}, J.~C., \& {Koss}, M.~J. 2021, \apj,
  907, 71, \dodoi{10.3847/1538-4357/abce5d}

\bibitem[{{Foord} {et~al.}(2019){Foord}, {G{\"u}ltekin}, {Reynolds},
  {Hodges-Kluck}, {Cackett}, {Comerford}, {King}, {Miller}, \&
  {Runnoe}}]{Foord+2019}
{Foord}, A., {G{\"u}ltekin}, K., {Reynolds}, M.~T., {et~al.} 2019, \apj, 877,
  17, \dodoi{10.3847/1538-4357/ab18a3}

\bibitem[{{Gao} {et~al.}(2003){Gao}, {Wang}, {Appleton}, \& {Lucas}}]{Gao+2003}
{Gao}, Y., {Wang}, Q.~D., {Appleton}, P.~N., \& {Lucas}, R.~A. 2003, \apjl,
  596, L171, \dodoi{10.1086/379598}

\bibitem[{{Greene} \& {Ho}(2007)}]{Greene&Ho2007}
{Greene}, J.~E., \& {Ho}, L.~C. 2007, \apj, 656, 84, \dodoi{10.1086/509064}

\bibitem[{{Guo} {et~al.}(2020){Guo}, {Inayoshi}, {Michiyama}, \&
  {Ho}}]{Guo+2020}
{Guo}, M., {Inayoshi}, K., {Michiyama}, T., \& {Ho}, L.~C. 2020, \apj, 901, 39,
  \dodoi{10.3847/1538-4357/abacc1}

\bibitem[{{Hirschmann} {et~al.}(2014){Hirschmann}, {Dolag}, {Saro}, {Bachmann},
  {Borgani}, \& {Burkert}}]{Hirschmann+2014}
{Hirschmann}, M., {Dolag}, K., {Saro}, A., {et~al.} 2014, \mnras, 442, 2304,
  \dodoi{10.1093/mnras/stu1023}

\bibitem[{{Hobbs} {et~al.}(2010){Hobbs}, {Archibald}, {Arzoumanian}, {Backer},
  {Bailes}, {Bhat}, {Burgay}, {Burke-Spolaor}, {Champion}, {Cognard}, {Coles},
  {Cordes}, {Demorest}, {Desvignes}, {Ferdman}, {Finn}, {Freire}, {Gonzalez},
  {Hessels}, {Hotan}, {Janssen}, {Jenet}, {Jessner}, {Jordan}, {Kaspi},
  {Kramer}, {Kondratiev}, {Lazio}, {Lazaridis}, {Lee}, {Levin}, {Lommen},
  {Lorimer}, {Lynch}, {Lyne}, {Manchester}, {McLaughlin}, {Nice}, {Oslowski},
  {Pilia}, {Possenti}, {Purver}, {Ransom}, {Reynolds}, {Sanidas}, {Sarkissian},
  {Sesana}, {Shannon}, {Siemens}, {Stairs}, {Stappers}, {Stinebring},
  {Theureau}, {van Haasteren}, {van Straten}, {Verbiest}, {Yardley}, \&
  {You}}]{Hobbs+2010}
{Hobbs}, G., {Archibald}, A., {Arzoumanian}, Z., {et~al.} 2010, Classical and
  Quantum Gravity, 27, 084013, \dodoi{10.1088/0264-9381/27/8/084013}

\bibitem[{{Holley-Bockelmann} {et~al.}(2008){Holley-Bockelmann},
  {G{\"u}ltekin}, {Shoemaker}, \& {Yunes}}]{Holley-Bockelmann+2008}
{Holley-Bockelmann}, K., {G{\"u}ltekin}, K., {Shoemaker}, D., \& {Yunes}, N.
  2008, \apj, 686, 829, \dodoi{10.1086/591218}

\bibitem[{{Holley-Bockelmann} \& {Khan}(2015)}]{Holley-Bockelmann&Khan2015}
{Holley-Bockelmann}, K., \& {Khan}, F.~M. 2015, \apj, 810, 139,
  \dodoi{10.1088/0004-637X/810/2/139}

\bibitem[{{Holley-Bockelmann} {et~al.}(2010){Holley-Bockelmann}, {Micic},
  {Sigurdsson}, \& {Rubbo}}]{Holley-Bockelmann+2010}
{Holley-Bockelmann}, K., {Micic}, M., {Sigurdsson}, S., \& {Rubbo}, L.~J. 2010,
  \apj, 713, 1016, \dodoi{10.1088/0004-637X/713/2/1016}

\bibitem[{{Hopkins} {et~al.}(2007){Hopkins}, {Richards}, \&
  {Hernquist}}]{Hopkins+2007}
{Hopkins}, P.~F., {Richards}, G.~T., \& {Hernquist}, L. 2007, \apj, 654, 731,
  \dodoi{10.1086/509629}

\bibitem[{{Igumenshchev} \& {Narayan}(2002)}]{Igumenshchev&Narayan2002}
{Igumenshchev}, I.~V., \& {Narayan}, R. 2002, \apj, 566, 137,
  \dodoi{10.1086/338077}

\bibitem[{{Izquierdo-Villalba} {et~al.}(2020){Izquierdo-Villalba}, {Bonoli},
  {Dotti}, {Sesana}, {Rosas-Guevara}, \& {Spinoso}}]{Izquierdo-Villalba+2020}
{Izquierdo-Villalba}, D., {Bonoli}, S., {Dotti}, M., {et~al.} 2020, \mnras,
  495, 4681, \dodoi{10.1093/mnras/staa1399}

\bibitem[{{Kaaret} {et~al.}(2001){Kaaret}, {Prestwich}, {Zezas}, {Murray},
  {Kim}, {Kilgard}, {Schlegel}, \& {Ward}}]{Kaaret+2001}
{Kaaret}, P., {Prestwich}, A.~H., {Zezas}, A., {et~al.} 2001, \mnras, 321, L29,
  \dodoi{10.1046/j.1365-8711.2001.04064.x}

\bibitem[{{King} \& {Dehnen}(2005)}]{King&Dehnen2005}
{King}, A.~R., \& {Dehnen}, W. 2005, \mnras, 357, 275,
  \dodoi{10.1111/j.1365-2966.2005.08634.x}

\bibitem[{{Knollmann} \& {Knebe}(2009)}]{Knollmann&Knebe2009}
{Knollmann}, S.~R., \& {Knebe}, A. 2009, \apjs, 182, 608,
  \dodoi{10.1088/0067-0049/182/2/608}

\bibitem[{{Kormendy} \& {Ho}(2013)}]{Kormendy&Ho2013}
{Kormendy}, J., \& {Ho}, L.~C. 2013, \araa, 51, 511,
  \dodoi{10.1146/annurev-astro-082708-101811}

\bibitem[{{Koss} {et~al.}(2012){Koss}, {Mushotzky}, {Treister}, {Veilleux},
  {Vasudevan}, \& {Trippe}}]{Koss+2012}
{Koss}, M., {Mushotzky}, R., {Treister}, E., {et~al.} 2012, \apjl, 746, L22,
  \dodoi{10.1088/2041-8205/746/2/L22}

\bibitem[{{Lehmer} {et~al.}(2016){Lehmer}, {Basu-Zych}, {Mineo}, {Brandt},
  {Eufrasio}, {Fragos}, {Hornschemeier}, {Luo}, {Xue}, {Bauer}, {Gilfanov},
  {Ranalli}, {Schneider}, {Shemmer}, {Tozzi}, {Trump}, {Vignali}, {Wang},
  {Yukita}, \& {Zezas}}]{Lehmer+2016}
{Lehmer}, B.~D., {Basu-Zych}, A.~R., {Mineo}, S., {et~al.} 2016, \apj, 825, 7,
  \dodoi{10.3847/0004-637X/825/1/7}

\bibitem[{{Libeskind} {et~al.}(2006){Libeskind}, {Cole}, {Frenk}, \&
  {Helly}}]{Libeskind+2006}
{Libeskind}, N.~I., {Cole}, S., {Frenk}, C.~S., \& {Helly}, J.~C. 2006, \mnras,
  368, 1381, \dodoi{10.1111/j.1365-2966.2006.10209.x}

\bibitem[{{Lin} {et~al.}(2018){Lin}, {Strader}, {Carrasco}, {Page},
  {Romanowsky}, {Homan}, {Irwin}, {Remillard}, {Godet}, {Webb}, {Baumgardt},
  {Wijnands}, {Barret}, {Duc}, {Brodie}, \& {Gwyn}}]{Lin+2018}
{Lin}, D., {Strader}, J., {Carrasco}, E.~R., {et~al.} 2018, Nature Astronomy,
  2, 656, \dodoi{10.1038/s41550-018-0493-1}

\bibitem[{{Lodato} \& {Natarajan}(2007)}]{Lodato&Natarajan2007}
{Lodato}, G., \& {Natarajan}, P. 2007, \mnras, 377, L64,
  \dodoi{10.1111/j.1745-3933.2007.00304.x}

\bibitem[{{Ma} {et~al.}(2021){Ma}, {Hopkins}, {Ma}, {Angl{\'e}s-Alc{\'a}zar},
  {Faucher-Gigu{\`e}re}, \& {Kelley}}]{Ma+2021}
{Ma}, L., {Hopkins}, P.~F., {Ma}, X., {et~al.} 2021, arXiv e-prints,
  arXiv:2101.02727.
\newblock \doarXiv{2101.02727}

\bibitem[{{Matsumoto} {et~al.}(2001){Matsumoto}, {Tsuru}, {Koyama}, {Awaki},
  {Canizares}, {Kawai}, {Matsushita}, \& {Kawabe}}]{Matsumoto+2001}
{Matsumoto}, H., {Tsuru}, T.~G., {Koyama}, K., {et~al.} 2001, \apjl, 547, L25,
  \dodoi{10.1086/318878}

\bibitem[{{Mezcua} {et~al.}(2018){Mezcua}, {Civano}, {Marchesi}, {Suh},
  {Fabbiano}, \& {Volonteri}}]{Mezcua+2018}
{Mezcua}, M., {Civano}, F., {Marchesi}, S., {et~al.} 2018, \mnras, 478, 2576,
  \dodoi{10.1093/mnras/sty1163}

\bibitem[{{Mezcua} \& {Dom{\'\i}nguez S{\'a}nchez}(2020)}]{Mezcua+2020}
{Mezcua}, M., \& {Dom{\'\i}nguez S{\'a}nchez}, H. 2020, \apjl, 898, L30,
  \dodoi{10.3847/2041-8213/aba199}

\bibitem[{{Miller} {et~al.}(2003){Miller}, {Fabbiano}, {Miller}, \&
  {Fabian}}]{Miller+2003}
{Miller}, J.~M., {Fabbiano}, G., {Miller}, M.~C., \& {Fabian}, A.~C. 2003,
  \apjl, 585, L37, \dodoi{10.1086/368373}

\bibitem[{{Natarajan}(2021)}]{Natarajan2021}
{Natarajan}, P. 2021, \mnras, 501, 1413, \dodoi{10.1093/mnras/staa3724}

\bibitem[{{Neumayer} {et~al.}(2020){Neumayer}, {Seth}, \&
  {B{\"o}ker}}]{Neumayer+2020}
{Neumayer}, N., {Seth}, A., \& {B{\"o}ker}, T. 2020, \aapr, 28, 4,
  \dodoi{10.1007/s00159-020-00125-0}

\bibitem[{{Pellegrini}(2005)}]{Pellegrini2005}
{Pellegrini}, S. 2005, \apj, 624, 155, \dodoi{10.1086/429267}

\bibitem[{{Perna} {et~al.}(2003){Perna}, {Narayan}, {Rybicki}, {Stella}, \&
  {Treves}}]{Perna+2003}
{Perna}, R., {Narayan}, R., {Rybicki}, G., {Stella}, L., \& {Treves}, A. 2003,
  \apj, 594, 936, \dodoi{10.1086/377091}

\bibitem[{{Pesce} {et~al.}(2021){Pesce}, {Seth}, {Greene}, {Braatz}, {Condon},
  {Kent}, \& {Krajnovi{\'c}}}]{Pesce+2021}
{Pesce}, D.~W., {Seth}, A.~C., {Greene}, J.~E., {et~al.} 2021, arXiv e-prints,
  arXiv:2101.07932.
\newblock \doarXiv{2101.07932}

\bibitem[{{Pfister} {et~al.}(2019){Pfister}, {Volonteri}, {Dubois}, {Dotti}, \&
  {Colpi}}]{Pfister+2019}
{Pfister}, H., {Volonteri}, M., {Dubois}, Y., {Dotti}, M., \& {Colpi}, M. 2019,
  \mnras, 486, 101, \dodoi{10.1093/mnras/stz822}

\bibitem[{{Pontzen} {et~al.}(2013){Pontzen}, {Ro{\v{s}}kar}, {Stinson}, \&
  {Woods}}]{Pontzen+2013}
{Pontzen}, A., {Ro{\v{s}}kar}, R., {Stinson}, G., \& {Woods}, R. 2013,
  {pynbody: N-Body/SPH analysis for python}.
\newblock \doeprint{1305.002}

\bibitem[{{Pontzen} \& {Tremmel}(2018)}]{Pontzen&Tremmel2018}
{Pontzen}, A., \& {Tremmel}, M. 2018, \apjs, 237, 23,
  \dodoi{10.3847/1538-4365/aac832}

\bibitem[{{Power} {et~al.}(2003){Power}, {Navarro}, {Jenkins}, {Frenk},
  {White}, {Springel}, {Stadel}, \& {Quinn}}]{Power+2003}
{Power}, C., {Navarro}, J.~F., {Jenkins}, A., {et~al.} 2003, \mnras, 338, 14,
  \dodoi{10.1046/j.1365-8711.2003.05925.x}

\bibitem[{{Rees}(1988)}]{Rees1988}
{Rees}, M.~J. 1988, \nat, 333, 523, \dodoi{10.1038/333523a0}

\bibitem[{{Reines} {et~al.}(2020){Reines}, {Condon}, {Darling}, \&
  {Greene}}]{Reines+2020}
{Reines}, A.~E., {Condon}, J.~J., {Darling}, J., \& {Greene}, J.~E. 2020, \apj,
  888, 36, \dodoi{10.3847/1538-4357/ab4999}

\bibitem[{{Reines} \& {Volonteri}(2015)}]{Reines&Volonteri2015}
{Reines}, A.~E., \& {Volonteri}, M. 2015, \apj, 813, 82,
  \dodoi{10.1088/0004-637X/813/2/82}

\bibitem[{{Ressler} {et~al.}(2021){Ressler}, {Quataert}, {White}, \&
  {Blaes}}]{Ressler+2021}
{Ressler}, S.~M., {Quataert}, E., {White}, C.~J., \& {Blaes}, O. 2021, \mnras,
  \dodoi{10.1093/mnras/stab311}

\bibitem[{{Ricarte} {et~al.}(2019){Ricarte}, {Tremmel}, {Natarajan}, \&
  {Quinn}}]{Ricarte+2019}
{Ricarte}, A., {Tremmel}, M., {Natarajan}, P., \& {Quinn}, T. 2019, \mnras,
  489, 802, \dodoi{10.1093/mnras/stz2161}

\bibitem[{{Ricarte} {et~al.}(2021){Ricarte}, {Tremmel}, {Natarajan}, {Zimmer},
  \& {Quinn}}]{Ricarte+2021}
{Ricarte}, A., {Tremmel}, M., {Natarajan}, P., {Zimmer}, C., \& {Quinn}, T.
  2021, arXiv e-prints, arXiv:2103.12124.
\newblock \doarXiv{2103.12124}

\bibitem[{{Sanchez} {et~al.}(2019){Sanchez}, {Werk}, {Tremmel}, {Pontzen},
  {Christensen}, {Quinn}, \& {Cruz}}]{Sanchez+2019}
{Sanchez}, N.~N., {Werk}, J.~K., {Tremmel}, M., {et~al.} 2019, \apj, 882, 8,
  \dodoi{10.3847/1538-4357/ab3045}

\bibitem[{{Schramm} \& {Silverman}(2013)}]{Schramm&Silverman2013}
{Schramm}, M., \& {Silverman}, J.~D. 2013, \apj, 767, 13,
  \dodoi{10.1088/0004-637X/767/1/13}

\bibitem[{{Steinborn} {et~al.}(2016){Steinborn}, {Dolag}, {Comerford},
  {Hirschmann}, {Remus}, \& {Teklu}}]{Steinborn+2016}
{Steinborn}, L.~K., {Dolag}, K., {Comerford}, J.~M., {et~al.} 2016, \mnras,
  458, 1013, \dodoi{10.1093/mnras/stw316}

\bibitem[{{Stemo} {et~al.}(2020){Stemo}, {Comerford}, {Barrows}, {Stern},
  {Assef}, {Griffith}, \& {Schechter}}]{Stemo+2020}
{Stemo}, A., {Comerford}, J.~M., {Barrows}, R.~S., {et~al.} 2020, arXiv
  e-prints, arXiv:2011.10051.
\newblock \doarXiv{2011.10051}

\bibitem[{{Stone} \& {Metzger}(2016)}]{Stone&Metzger2016}
{Stone}, N.~C., \& {Metzger}, B.~D. 2016, \mnras, 455, 859,
  \dodoi{10.1093/mnras/stv2281}

\bibitem[{{Stone} {et~al.}(2020){Stone}, {Vasiliev}, {Kesden}, {Rossi},
  {Perets}, \& {Amaro-Seoane}}]{Stone+2020}
{Stone}, N.~C., {Vasiliev}, E., {Kesden}, M., {et~al.} 2020, \ssr, 216, 35,
  \dodoi{10.1007/s11214-020-00651-4}

\bibitem[{{Strubbe} \& {Quataert}(2009)}]{Strubbe&Quataert2009}
{Strubbe}, L.~E., \& {Quataert}, E. 2009, \mnras, 400, 2070,
  \dodoi{10.1111/j.1365-2966.2009.15599.x}

\bibitem[{{Tamfal} {et~al.}(2018){Tamfal}, {Capelo}, {Kazantzidis}, {Mayer},
  {Potter}, {Stadel}, \& {Widrow}}]{Tamfal+2018}
{Tamfal}, T., {Capelo}, P.~R., {Kazantzidis}, S., {et~al.} 2018, \apjl, 864,
  L19, \dodoi{10.3847/2041-8213/aada4b}

\bibitem[{{Tremmel} {et~al.}(2018{\natexlab{a}}){Tremmel}, {Governato},
  {Volonteri}, {Pontzen}, \& {Quinn}}]{Tremmel+2018a}
{Tremmel}, M., {Governato}, F., {Volonteri}, M., {Pontzen}, A., \& {Quinn},
  T.~R. 2018{\natexlab{a}}, \apjl, 857, L22, \dodoi{10.3847/2041-8213/aabc0a}

\bibitem[{{Tremmel} {et~al.}(2015){Tremmel}, {Governato}, {Volonteri}, \&
  {Quinn}}]{Tremmel+2015}
{Tremmel}, M., {Governato}, F., {Volonteri}, M., \& {Quinn}, T.~R. 2015,
  \mnras, 451, 1868, \dodoi{10.1093/mnras/stv1060}

\bibitem[{{Tremmel} {et~al.}(2018{\natexlab{b}}){Tremmel}, {Governato},
  {Volonteri}, {Quinn}, \& {Pontzen}}]{Tremmel+2018b}
{Tremmel}, M., {Governato}, F., {Volonteri}, M., {Quinn}, T.~R., \& {Pontzen},
  A. 2018{\natexlab{b}}, \mnras, 475, 4967, \dodoi{10.1093/mnras/sty139}

\bibitem[{{Tremmel} {et~al.}(2017){Tremmel}, {Karcher}, {Governato},
  {Volonteri}, {Quinn}, {Pontzen}, {Anderson}, \& {Bellovary}}]{Tremmel+2017}
{Tremmel}, M., {Karcher}, M., {Governato}, F., {et~al.} 2017, \mnras, 470,
  1121, \dodoi{10.1093/mnras/stx1160}

\bibitem[{{Tremmel} {et~al.}(2019){Tremmel}, {Quinn}, {Ricarte}, {Babul},
  {Chadayammuri}, {Natarajan}, {Nagai}, {Pontzen}, \&
  {Volonteri}}]{Tremmel+2019}
{Tremmel}, M., {Quinn}, T.~R., {Ricarte}, A., {et~al.} 2019, \mnras, 483, 3336,
  \dodoi{10.1093/mnras/sty3336}

\bibitem[{{van den Bosch} \& {Ogiya}(2018)}]{vandenBosch+2018}
{van den Bosch}, F.~C., \& {Ogiya}, G. 2018, \mnras, 475, 4066,
  \dodoi{10.1093/mnras/sty084}

\bibitem[{{van Velzen} {et~al.}(2011){van Velzen}, {Farrar}, {Gezari},
  {Morrell}, {Zaritsky}, {{\"O}stman}, {Smith}, {Gelfand}, \&
  {Drake}}]{vanVelzen+2011}
{van Velzen}, S., {Farrar}, G.~R., {Gezari}, S., {et~al.} 2011, \apj, 741, 73,
  \dodoi{10.1088/0004-637X/741/2/73}

\bibitem[{{Volonteri}(2007)}]{Volonteri2007}
{Volonteri}, M. 2007, \apjl, 663, L5, \dodoi{10.1086/519525}

\bibitem[{{Volonteri} {et~al.}(2016){Volonteri}, {Dubois}, {Pichon}, \&
  {Devriendt}}]{Volonteri+2016}
{Volonteri}, M., {Dubois}, Y., {Pichon}, C., \& {Devriendt}, J. 2016, \mnras,
  460, 2979, \dodoi{10.1093/mnras/stw1123}

\bibitem[{{Volonteri} {et~al.}(2003){Volonteri}, {Haardt}, \&
  {Madau}}]{Volonteri+2003}
{Volonteri}, M., {Haardt}, F., \& {Madau}, P. 2003, \apj, 582, 559,
  \dodoi{10.1086/344675}

\bibitem[{{Volonteri} \& {Perna}(2005)}]{Volonteri&Perna2005}
{Volonteri}, M., \& {Perna}, R. 2005, \mnras, 358, 913,
  \dodoi{10.1111/j.1365-2966.2005.08832.x}

\bibitem[{{Volonteri} {et~al.}(2020){Volonteri}, {Pfister}, {Beckmann},
  {Dubois}, {Colpi}, {Conselice}, {Dotti}, {Martin}, {Jackson}, {Kraljic},
  {Pichon}, {Trebitsch}, {Yi}, {Devriendt}, \& {Peirani}}]{Volonteri+2020}
{Volonteri}, M., {Pfister}, H., {Beckmann}, R.~S., {et~al.} 2020, \mnras, 498,
  2219, \dodoi{10.1093/mnras/staa2384}

\end{thebibliography}

\appendix

\section{Raw Versus Accreted Black Hole Masses in {\sc Romulus}}
\label{sec:raw_accreted_mass}

\begin{figure*}
  \centering
  \includegraphics[width=\textwidth]{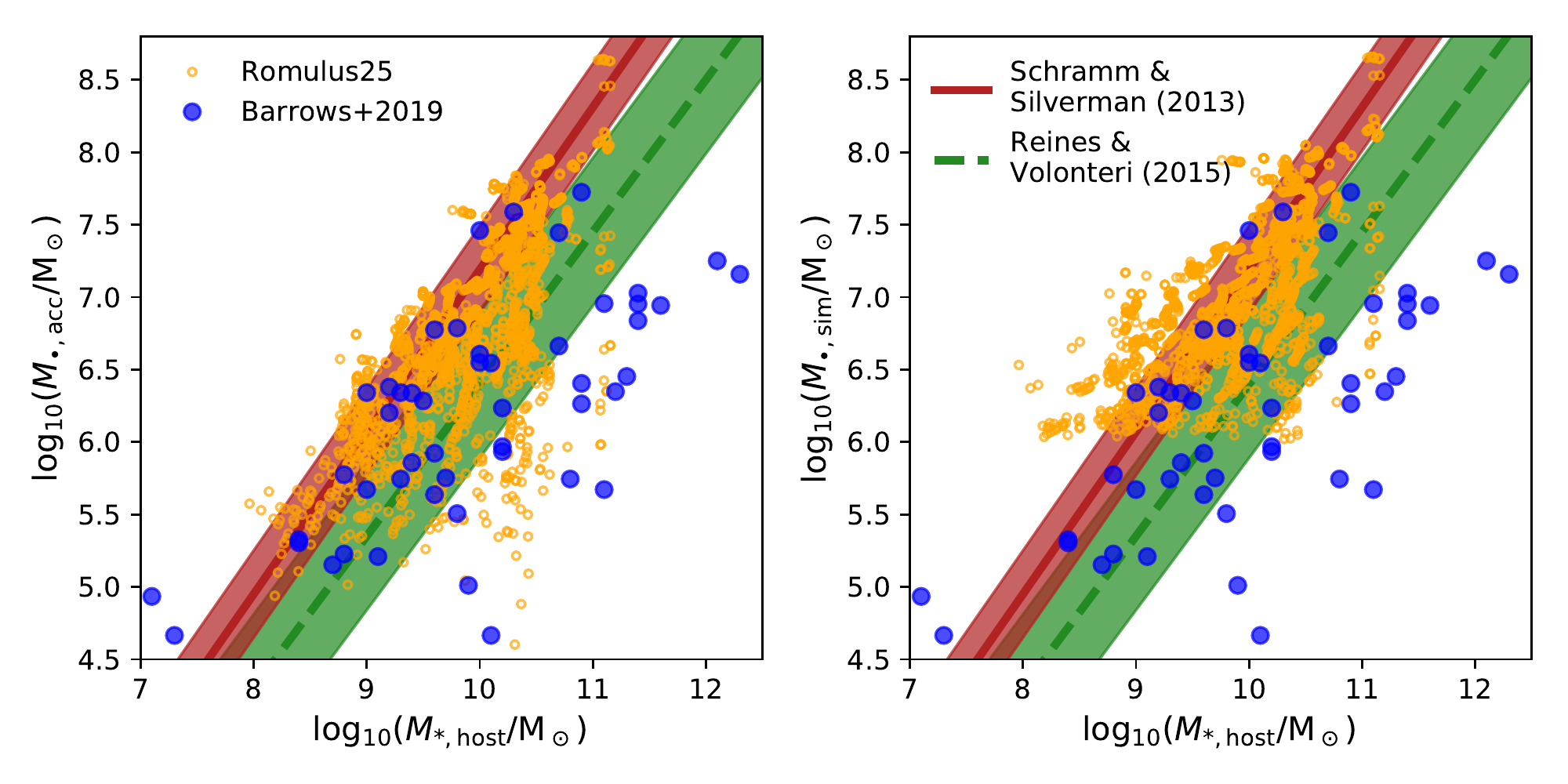}
  \caption{Here, we compare accreted ({\it left}) and raw ({\it right}) SMBH masses in the {\sc Romulus} simulations in the context of Figure \ref{fig:barrows_host_comparison}.  As pointed out in \citet{Ricarte+2019}, the accreted mass follows the $M_\bullet-M_*$ relation to which this simulation is calibrated even below the seed mass of $10^6 \ \mathrm{M}_\odot$, making it a valuable proxy for the true SMBH mass in low-mass galaxies.  If raw masses are used instead, as in the right panel, SMBH masses are all pushed above the seed mass, but we retain general agreement with the \citet{Schramm&Silverman2013} relationship. \label{fig:barrows_host_comparison_accreted_mass_check}}
\end{figure*}

In Figure \ref{fig:barrows_host_comparison}, we compared the $M_\bullet-M_*$ relation observed for HLXs with subsequently detected host galaxies with that from the {\sc Romulus} simulations.  For this comparison, we used only the accreted portion of a SMBH's mass, excluding seed masses.  This is because the accreted mass in the {\sc Romulus} simulations actually follows the $M_\bullet-M_*$ at low masses better than the raw mass, even below the seed mass of $10^6 \ \mathrm{M}_\odot$ \citep{Ricarte+2019}.  In Figure \ref{fig:barrows_host_comparison_accreted_mass_check}, we compare the accreted mass $M_{\bullet,\mathrm{acc}}$ with $M_{\bullet,\mathrm{sim}}$, the raw SMBH from the {\sc Romulus} simulations.  While lower masses are pushed to values above the seed mass, we find general agreement.

\section{Projected Flux Density Profiles}
\label{sec:xrbs}

Here, we detail how we estimate the X-ray contribution to projected flux density profiles by XRBs.  By analyzing the 6 Ms {\it Chandra} Deep Field South, \citet{Lehmer+2016} arrive at the following relationship between the 2-10 keV X-ray luminosity $L_x$, galaxy stellar mass $M_*$, star formation rate $\dot{M}_*$, and redshift $z$: 

\begin{equation}
  \left( \frac{L_x}{\mathrm{erg}\;\mathrm{s}^{-1}} \right) = \alpha_0(1+z)^\gamma \left( \frac{M_*}{\mathrm{M}_\odot} \right) + \beta_0(1+z)^\delta \left( \frac{\dot{M}_*}{\mathrm{M}_\odot \; \mathrm{yr}^{-1}} \right).
  \label{eqn:lehmer}
\end{equation}

\noindent where $\log \alpha_0 = 29.37 \pm 0.15$, $\log \beta_0 = 39.28 \pm 0.05$, $\gamma=2.03 \pm 0.60$, and $\delta = 1.31 \pm 0.13$.  In the {\sc Romulus} simulations, we compute radial profiles of the stellar mass and star formation rate density and average these together for halos in a given halo mass bin.  We apply Equation \ref{eqn:lehmer} to transform these into X-ray luminosity density profiles in units of $\mathrm{erg} \; \mathrm{s}^{-1} \; \mathrm{cm}^{-3}$.  An Abel transform is then required to turn this into a projected luminosity density profile $l_x(r)$ in units of $\mathrm{erg} \; \mathrm{s}^{-1} \; \mathrm{cm}^{-2}$.

The projected flux density profile is then obtained via $f_x = l_x / (4 \pi d_L^2)$, where $d_L$ is the luminosity distance.  Finally, we transform between physical scales ($r$) and angular scales ($\theta$) by computing the angular diameter distance, such that $\theta = r/d_A$.  The final projected flux density profile is given by

\begin{equation}
  \mathcal{F}(\theta(r)) \ [\mathrm{erg} \; \mathrm{s}^{-1} \; \mathrm{cm}^{-2} \; \mathrm{deg}^{-2}]= \frac{l_x(r(\theta)) d_A^2}{4 \pi d_L^2} \left( \frac{\pi \ \mathrm{rad}}{180 \ \mathrm{deg}} \right)^2.
\end{equation}

\end{document}